\documentclass[aps,prd,preprintnumbers,superscriptaddress,nofootinbib]{revtex4}

\usepackage{lmodern}
\usepackage{hyperref}
\hypersetup{colorlinks=true,linkcolor=purple,anchorcolor=blue,citecolor=blue, filecolor=blue,urlcolor=blue,bookmarksnumbered=true,
pdfview=FitB
}
\usepackage{mathrsfs}
\usepackage{rsfso}
\usepackage{color}
\usepackage{xcolor}
\usepackage{ulem}
\usepackage{csquotes}
\colorlet{purple1}{blue!70!red}
\colorlet{darkred}{red!50!black}

\usepackage{graphicx}
\usepackage[bf,SL,BF]{subfigure}
\usepackage{psfrag}
\usepackage{color}
\usepackage{amssymb}
\usepackage{amsmath}
\usepackage{epstopdf}
\usepackage{natbib}
\usepackage{amssymb}
\usepackage{amsmath,amssymb}
\usepackage{mathtools}
\usepackage{float}
\usepackage{color}
\usepackage{leftidx}
\usepackage{booktabs}
\usepackage{latexsym}
\usepackage{revsymb}
\usepackage{multirow}
\usepackage{hypcap}
\usepackage{array}
\usepackage[english]{babel}
\usepackage{amsmath}
\usepackage{epstopdf}

\newcommand{\nslash}{\kern 0.2 em n\kern -0.50em /}
\newcommand{\kslash}{\kern 0.2 em k\kern -0.45em /}
\newcommand{\lslash}{\kern 0.2 em l\kern -0.50em /}
\newcommand{\pslash}{\kern 0.2 em p\kern -0.50em /}
\newcommand{\Sslash}{\kern 0.2 em S\kern -0.50em /}
\newcommand{\Pslash}{\kern 0.2 em P\kern -0.50em /}
\newcommand{\Dslash}{\kern 0.2 em D\kern -0.65em /\kern 0.15em}

\newcommand{\Sop}{\boldsymbol{\hat S}}

\newcommand{\be}{\begin{eqnarray}}
	\newcommand{\ee}{\end{eqnarray}}

\newcommand{\bfp}{{\bf p}_{\perp}}

\newcommand{\CM}{\color{purple}}

\begin{document}
	
	\title{Gluon distributions in the proton in a light-front spectator model} 

 	\author{Dipankar~Chakrabarti}
	\email{dipankar@iitk.ac.in} 
	\affiliation{Department of Physics, Indian Institute of Technology Kanpur, Kanpur-208016, India}
 
    \author{Poonam~Choudhary}
    \email{poonamch@iitk.ac.in} 
    \affiliation{Department of Physics, Indian Institute of Technology Kanpur, Kanpur-208016, India}

	\author{Bheemsehan~Gurjar}
	\email{gbheem@iitk.ac.in} 
	\affiliation{Department of Physics, Indian Institute of Technology Kanpur, Kanpur-208016, India}

 \author{Raj Kishore}
	\email{raj.theps@gmail.com} 
	\affiliation{Department of Physics, Indian Institute of Technology Kanpur, Kanpur-208016, India}

  \author{Tanmay Maji}
  \email{tanmayhep@gmail.com}
  \affiliation{Department of Physics, National Institute of Technology Kurukshetra, Kurukshetra-136119, India}

  \author{Chandan Mondal}
  \email{mondal@impcas.ac.cn}
  \affiliation{Institute of Modern Physics, Chinese Academy of Sciences, Lanzhou 730000, China}
  \affiliation{School of Nuclear Science and Technology, University of Chinese Academy of Sciences, Beijing 100049, China}
	
 \author{Asmita~Mukherjee}
	\email{asmita@phy.iitb.ac.in} 
	\affiliation{Department of Physics, Indian Institute of Technology Bombay, Powai, Mumbai 400076, India}
	

	\begin{abstract}
  We formulate a light-front spectator model for the proton incorporating the gluonic degree of freedom. In this model, at high energy scattering of the proton, 
 the active parton is a gluon and the rest is viewed as a spin-$\frac{1}{2}$ spectator with an effective mass.
  The light front wave functions of the proton are constructed using a soft wall AdS/QCD prediction and parameterized by fitting the unpolarized gluon distribution function to the NNPDF3.0nlo data set.
 We investigate the helicity 
 distribution of gluon in this model. We find that our prediction for the gluon helicity asymmetry agrees well with existing
experimental data and satisfies the perturbative QCD constraints at small and large longitudinal momentum regions.  We also present the transverse momentum dependent distributions (TMDs) for gluon in this model. We further show that the model-independent Mulders-Rodrigues inequalities are obeyed by the TMDs computed in our model. 
	\end{abstract}
 
	\maketitle
	\section{Introduction}\label{sec:intro}
	Understanding the structure of hadrons in terms of the fundamental degrees of freedom in QCD i.e., quarks and gluons, is one of the remaining challenges in nuclear and particle physics. There have been numerous research in recent years to learn more about the parton distributions~(PDFs), transverse momentum dependent distributions (TMDs), generalized parton distributions~(GPDs), gravitational form factors~(GFFs), Wigner distributions, etc., of the quarks and their properties were investigated using various theoretical models~\cite{Meissner:2007rx,Jakob:1997wg,Brodsky:2002cx,Bacchetta:2008af,Pasquini:2008ax,Maji:2017bcz,Gurjar:2021dyv,Gurjar:2022rcl,Lorce:2011zta,Ji:1997gm,Scopetta:2002xq,Boffi:2002yy,Boffi:2003yj,Vega:2010ns,Chakrabarti:2013gra,Mondal:2015uha,Chakrabarti:2015ama,Xu:2021wwj,Kriesten:2021sqc,Chakrabarti:2016yuw,Meissner:2008ay,Meissner:2009ww,Burkert:2023wzr}, revealing numerous  insights into the nucleon structure. In comparison to quark distributions, the gluon distributions are less precisely known, which  has an impact on the calculation of the cross-section of a process dominated by the 
    gluon-initiated channel. Gluons, which mediate the strong interaction, play a crucial role in the mass and spin of the nucleon~\cite{Ji:2020ena,Jaffe:1989jz,PhysRevLett.78.610,Leader:2021pqf,Liu:2015xha}. In the study of deep inelastic scattering processes, the gluon distributions and fragmentation functions contain essential information about the system~\cite{Xie:2022lra}. These process-independent distributions characterize the soft part of the scattering, or the deep structure of the hadrons, together with their quark and antiquark equivalents.
	The majority of hadron high energy scattering investigations relies on the QCD factorization, in which PDFs play a crucial role. There are two gluonic PDFs at leading twist: unpolarized $f_1^g(x)$ and polarized $g_{1L}^g(x)$ PDFs. The unpolarized gluon PDF has been studied using  lattice QCD~\cite{HadStruc:2021wmh,Yang:2016plb,Delmar:2022plq,Sufian:2020wcv} and various other theoretical approaches~\cite{Brodsky:2000ii,Dosch:2022mop,Hou:2019efy,NNPDF:2017mvq,Delmar:2022plq,Freese:2021zne} with better accuracy as compared to the polarized PDF. The uncertainty is mainly in the small $x$ region of the polarized PDF. Even the sign of the polarized PDF in the small $x$ region is not yet well-decided~\cite{deFlorian:2014yva}.
	
The first Mellin moment of the gluon polarized PDF gives the gluon spin contribution to the proton spin. It is found that the quark spin only contributes around $(20 - 30)\%$ of the proton spin~\cite{EuropeanMuon:1987isl,EuropeanMuon:1989yki,deFlorian:2009vb,Nocera:2014gqa,Ethier:2017zbq}. Even after separating out the quark OAM and spin contributions, there is a sizable amount of  spin contribution that can not be explained through quarks only.
 Several experiments like the RHIC spin program
at BNL~\cite{Nocera:2014gqa,Ethier:2017zbq,deFlorian:2014yva}, PHENIX~\cite{PHENIX:2014gbf,PHENIX:2008swq} and COMPASS~\cite{COMPASS:2015mhb} observed a non-zero gluon helicity suggesting that the proton spin is significantly influenced by the gluon, which is important to resolve the proton spin puzzle. However, the gluon helicity is not well determined yet. The polarized PDF $g_{1L}^g(x)$ measures the gluon spin contribution to the proton but it has large uncertainty in the small-$x$ region. Recently, some theoretical studies using the  basis light-front quantization (BLFQ) approach~\cite{Xu:2022abw}, the holographic light-front QCD (HLFQCD) approach~\cite{Gurjar:2022jkx}, and the lattice QCD~\cite{Khan:2022vot,HadStruc:2022yaw,Sufian:2020wcv}  have reported the nonzero and sizeable contributions of the gluon spin to the total proton spin. 
While there have been significant improvements in the extracted $g_{1L}^g(x)$ precision over the last decade~\cite{Gehrmann:1995ag,Gluck:2000dy,Blumlein:2002qeu}, there are still various concerns, such as the suppression in the gluon distribution in the momentum fraction region $ 0.1 < x<0.4 $ when the ATLAS and CMS jet data are included \cite{Hou:2019efy}, determination of gluon helicity in entire $x$-region etc. 
Accurate measurement of the nucleon spin structure, specifically the gluon and sea quark distributions, are two of the primary scientific objectives of the forthcoming Electron-Ion-Colliders (EICs)~\cite{Accardi:2012qut,AbdulKhalek:2021gbh,Anderle:2021wcy}. It is not yet completely clear how gluons, valence quarks, and sea quarks share the mass and the spin of the proton. But in recent studies, it has been found that gluons contribute more to the proton spin than the valence quarks. 
The Electron-Ion-Collider (EIC) will focus particularly on the small-$x$ region, 
 which contains more gluon density and is complicated to study. 
 Also, the gluons play a crucial role in determining the mass of the proton, as their interactions with quarks contribute significantly to the overall mass of the proton~\cite{Ji:1994av,Ji:1995sv}. Understanding the role of gluons in proton mass decomposition is an active area of research in theoretical and experimental physics, and has important implications for our understanding of the fundamental building blocks of matter~\cite{Burkert:2023wzr,Rodini:2020pis, Metz:2020vxd,Lorce:2021xku}.

In addition to PDFs, we also explore the gluon TMDs in this work. It has recently been demonstrated that gluons and quark TMDs are essential to describe the three-dimensional picture of the nucleon in momentum space. The gluon TMDs have been studied in \cite{DAlesio:2018rnv,Mulders:2000sh,Boer:2015pni,Lyubovitskij:2021qza,Bacchetta:2020vty,Lu:2016vqu}. TMDs play a crucial role in the experimentally observed single spin and azimuthal asymmetries in for example semi-inclusive deep inelastic scattering (SIDIS) and Drell-Yan (DY) processes. There are eight leading twist gluon TMDs. The collinear limit of the TMDs $f_1^g(x,\bfp^2)$ and $g_{1L}^g(x,\bfp^2)$ are related to the unpolarized and polarized  PDFs, respectively. 
An overview of the available literature on unpolarized and helicity gluon TMDs at small-$x$ can be found in Ref.~\cite{Petreska:2018cbf} (and references therein). Some  recent theoretical and phenomenological studies are discussed in Refs.~\cite{Altinoluk:2018byz,Altinoluk:2020qet,Zhou:2018lfq}.


 Recently, a few spectator models have been proposed for the study of the gluon distributions \cite{Bacchetta:2020vty,Lyubovitskij:2020xqj,Lu:2016vqu,Tan:2023kbl}. In the construction of the spectator models, the crucial step is the choice of the light-front wave functions \cite{Brodsky:2000ii}. We can write the form factors, PDFs, TMDs, GPDs and Wigner distributions in terms of the light-front wave functions in a spectator model. It has been verified that the form factors and partonic distributions follow model-independent scaling rules in the limiting cases of the longitudinal momentum fraction $x$. The behaviour of gluon parton densities at large and small $x$ have been observed in Refs.~\cite{Brodsky:1989db,Brodsky:1994kg}. In these works, the authors derived QCD constraints ~\cite{Brodsky:1989db} on unpolarized $ f_{1}^{g}$, polarized  $g_{1L}^{g}$ gluon PDFs and on the gluon helicity asymmetry ratio $g_{1L}^{g}/f_{1}^{g}$, which goes to zero as $x \to 0$ and it is 1 as $x \to 1$ . A reasonably good model should follow these limiting conditions. In this work, we study the gluon PDFs and T-even gluon TMDs using a light-front gluon spectator model, where the light-front wave functions are constructed using a soft wall AdS/QCD prediction~\cite{Brodsky:2014yha}. This model is a generalization of the light-front quark-diquark model~\cite{Gutsche:2013zia,Maji:2016yqo,Mondal:2015uha}. The gluon-spectator model describes the nucleon as a composite system of an active gluon and the rest of the system as a spectator. At low energy, the spectator contains mainly three valence quarks of the nucleon. We are considering the spectator as a spin-$\frac{1}{2}$ effective system.

 
	 The paper is organized as follows: In Sec.~\ref{sec:model}, we discuss about our model construction.  We show the model calculation of the T-even TMDs in Sec.~\ref{sec:TMDs}. In Sec.~\ref{sec:fitting}, we determine our model parameters from the fitting of our unpolarized gluon PDF to the NNPDF3.0nlo data. In Sec.~\ref{sec:Results}, we show our model results for all four T-even gluon TMDs. Finally, we provide a {\CM brief}
	 summary and discussion in Sec.~\ref{sec:concl}.
  \\
  
	\section{Model construction }\label{sec:model}
	The minimum Fock state of the proton contains only valence quarks. As we include the higher Fock sectors then  gluons, and sea quarks also come into the picture.  In this simplified model, we describe the proton as a composite state of one active gluon and a spin-$\frac{1}{2}$ spectator \cite{Lu:2016vqu}. 
	
 We choose a reference frame in which the transverse momentum of the proton vanishes, i.e. $ P=(P^+,\frac{M^2}{P^+},\textbf{0}_\perp)$. The momentum of  the active parton  is given by $ p=(x P^+, \frac{p^2+\textbf{p}_\perp^2}{x P^+}, \textbf{p}_\perp)$ and the momentum of the spectator $ P_X=((1-x) P^+, P^-_X, -\textbf{p}_\perp)$ 
 with $x=p^{+}/P^{+}$ being the fraction of proton longitudinal momentum carried by the struck gluon.  The proton state can be written as a two-particle Fock-state expansion with proton spin components $J_{z}=\pm\frac{1}{2}$~\cite{Brodsky:2000ii} as,
	\begin{eqnarray}\label{state}\nonumber
		|P;\uparrow(\downarrow)\rangle
		= \int \frac{\mathrm{d}^2 \bfp \mathrm{d} x}{16 \pi^3 \sqrt{x(1-x)}}\times \Bigg[\psi_{+1+\frac{1}{2}}^{\uparrow(\downarrow)}\left(x, \bfp\right)\left|+1,+\frac{1}{2} ; x P^{+}, \bfp\right\rangle+\psi_{+1-\frac{1}{2}}^{\uparrow(\downarrow)}\left(x, \bfp \right)\left|+1,-\frac{1}{2} ; x P^{+}, \bfp \right\rangle\\ 
		+\psi_{-1+\frac{1}{2}}^{\uparrow(\downarrow)}\left(x, \bfp \right)\left|-1,+\frac{1}{2} ; x P^{+}, \bfp \right\rangle+\psi_{-1-\frac{1}{2}}^{\uparrow(\downarrow)}\left(x, \bfp\right)\left|-1,-\frac{1}{2} ; x P^{+}, \bfp\right\rangle\bigg],
	\end{eqnarray}
	where $\psi_{\lambda_{g}\lambda_{X}}^{\uparrow(\downarrow)}(x,\bfp)$ are the LFWFs  corresponding to the two-particle state $|\lambda_{g},\lambda_{X};xP^{+},\bfp \rangle$ with proton helicities $\lambda_p=\uparrow(\downarrow)$. Here, 
 $\lambda_{g}$ and $\lambda_{X}$ stand for the helicity components of the constituent gluon and spectator, respectively. 
	
	Our proposal for the light-front wave functions in Eq.~(\ref{state}) is inspired by the wave function of the physical electron ~\cite{Brodsky:2000ii}, which is made up of a spin-1 photon and a spin-$\frac{1}{2}$ electron. We argue that the light-front wave functions for the Fock-state expansion for a proton with $J_{z}=+1/2$ have the following form:
	\begin{eqnarray} \label{LFWFsuparrow}   \nonumber
		\psi_{+1+\frac{1}{2}}^{\uparrow}\left(x,\bfp\right)&=&-\sqrt{2}\frac{(-p^{1}_{\perp}+ip^{2}_{\perp})}{x(1-x)}\varphi(x,\bfp^2), \\ \nonumber
		\psi_{+1-\frac{1}{2}}^{\uparrow}\left(x, \bfp\right)&=&-\sqrt{2}\bigg( M-\frac{M_{X}}{(1-x)} \bigg) \varphi(x,\bfp^2), \\ \nonumber
		\psi_{-1+\frac{1}{2}}^{\uparrow}\left(x, \bfp\right)&=&-\sqrt{2}\frac{(p^{1}_{\perp}+ip^{2}_{\perp})}{x}\varphi(x,\bfp^2), \\
		\psi_{-1-\frac{1}{2}}^{\uparrow}\left(x, \bfp\right)&=&0,
	\end{eqnarray}
	where $M$ and $M_X$ represent the masses of the proton and spectator, respectively.  $\varphi(x,\bfp^2)$ is the modified form of the soft-wall AdS/QCD wave function~\cite{Gutsche:2013zia} modeled by introducing the parameters $a$ and $b$.
	Similarly, the light-front wave functions for the proton with $J_{z}=-1/2$  have the form
	\begin{eqnarray} \label{LFWFsdownarrow}   \nonumber
		\psi_{+1+\frac{1}{2}}^{\downarrow}\left(x, \bfp\right)&=& 0, \\ \nonumber
		\psi_{+1-\frac{1}{2}}^{\downarrow}\left(x,\bfp\right)&=&-\sqrt{2}\frac{(-p^{1}_{\perp}+ip^{2}_{\perp})}{x}\varphi(x,\bfp^2), \\ \nonumber
		\psi_{-1+\frac{1}{2}}^{\downarrow}\left(x, \bfp\right)&=&-\sqrt{2}\bigg( M-\frac{M_{X}}{(1-x)} \bigg) \varphi(x,\bfp^2),  \\
		\psi_{-1-\frac{1}{2}}^{\downarrow}\left(x, \bfp \right)&=& -\sqrt{2}\frac{(p^{1}_{\perp}+ip^{2}_{\perp})}{x(1-x)}\varphi(x,\bfp^2).
	\end{eqnarray}

 %
 The behaviour at $x \rightarrow 0$, as well as the counting rules at $x \rightarrow 1$, provide information on the various gluon distributions 
 ~\cite{Brodsky:1989db,Brodsky:1994kg}. To elaborate, the asymptotic behaviour of the PDFs at small $x$  is adopted from the observed Regge behaviour in particle colliders, and the large-$x$ behavior is based on the power counting rules for hard scattering \cite{Brodsky:1994kg}. Keeping all these in mind, we have modified the soft wall AdS-QCD wave function, $\varphi(x,\bfp^{2})$,
 The complete form of the modified soft-wall AdS/QCD wave function is given by,
	\be\label{AdSphi}
	\varphi(x,\bfp^2)=N_{g}\frac{4\pi}{\kappa}\sqrt{\frac{\log[1/(1-x)]}{x}}x^{b}(1-x)^{a}\exp{\bigg[-\frac{\log[1/(1-x)]}{2\kappa^{2}x^2}\bfp^{2}\bigg]}.
	\ee	
	where $a$ and $b$ are our model parameters. The values of the model parameters $a$, $b$, and the normalization constant $N_{g}$ are fixed by fitting the gluon unpolarized PDF at the scale $\mu_0=2$ GeV with NNPDF3.0 data. For the stability of the proton, the spectator mass, $M_{X}$ is considered higher than the proton mass i.e.,
 $M_X>M$. 


	\section{Gluon TMDs }\label{sec:TMDs}
	
	In the light front formalism, the unintegrated gluon correlation function for leading twist gluon TMDs in the SIDIS process is given by the following relation~\cite{Mulders:2000sh}: 
	\begin{eqnarray} \label{eq:correlator}
		\Phi^{g[ij]}(x,\bfp;S)
		=\frac{1}{xP^+}\int\frac{d\xi^-}{2\pi}\,\frac{d^2 \mathbf{\xi}{_\perp}}{(2\pi)^2}\,e^{i k\cdot \xi}\,
		\big<P;S\big|\,
		F^{+j}_a(0) \ \,\mathcal{W}_{+\infty,ab}(0;\xi)\,
		F^{+i}_b(\xi)\,\big|P;S\big>\,
		\Big|_{\xi^+=0^+} \,,
	\end{eqnarray}
	where $F^{\mu\nu}$ is the gluon field strength tensor and $\mathcal{W}_{+\infty, ab}$ is the Wilson line that  ensures the correlator to be gauge invariant. 
	The subscript ``$+$'' specifies that the Wilson line in the correlator operator expression is future-pointing, which is necessary for SIDIS TMD distributions.
	There are eight leading twist gluon TMDs out of which four of them are T-even ($f_{1}^{g}$, $g_{1L}^{g}$, $g_{1T}^{g}$, and $h_{1}^{\perp g}$) and the remaining four are T-odd ($f_{1T}^{\bot g},\;h_{1L}^{\bot g},\;h_{1T}^{g},\;h_{1T}^{\bot g}$)~\cite{Meissner:2007rx, Mulders:2000sh}.
	The twist-2 gluon TMDs are defined through the correlator~(\ref{eq:correlator}) as \cite{Meissner:2007rx}
	\begin{eqnarray} \label{eq:gtmd1}
		\Phi^g(x,\bfp;S)&=&\delta_T^{ij}\,\Phi^{g[ij]}(x,\bfp;S)\\ \nonumber
		&=&f_1^g(x,\bfp^{2})
		-\frac{\epsilon^{ij}_{\perp} \bfp^i S_\perp^j}{M}\,f_{1T}^{\bot g}(x,\bfp^{\,2}) \,,
	\end{eqnarray}
	
	\begin{eqnarray}\label{eq:gtmd2}
		\tilde\Phi^g(x,\bfp;S)&=&i\epsilon_T^{ij}\,\Phi^{g[ij]}(x,\bfp;S)\\ \nonumber
		&=&\lambda\,g_{1L}^g(x,\bfp^{\,2})
		+\frac{\bfp\cdot\mathbf{S}_{\perp}}{M}\,g_{1T}^g(x,\bfp^{\,2}) \,,
	\end{eqnarray}
	\begin{align}\label{eq:gtmd3}
		\Phi_T^{g,\,ij}(x,\bfp;S)&=-\Sop\,\Phi^{g[ij]}(x,\bfp;S)\\ \nonumber
		&=-\frac{\hat{\mathbf{S}} \bfp^i \bfp^j}{2M^2}\,h_{1}^{\perp g}(x,\bfp^{2})
		+\frac{\lambda\,\Sop \bfp^i\epsilon_{\perp}^{jk}\bfp^k}{2M^2}\,h_{1L}^{\perp g}(x,\bfp^{2})\\ \nonumber
		&+\frac{\hat{\bm{S}} \bfp^{i}\epsilon_{\perp}^{jk}S_{\perp}^{k}}{2M}\,\bigg(h_{1T}^g(x,\bfp^{\,2})+\frac{\bfp^{\,2}}{2M^2}\,h_{1T}^{\bot g}(x,\bfp^{\,2})\bigg)\\ \nonumber
		&+\frac{\hat{\bm{S}} \bfp^{i}\epsilon_{\perp}^{jk}\big(2p_{\perp}^k \bfp.\mathbf{S}_{\perp}-S_{\perp}^k \bfp^{\,2}\big)}{4M^{3}}h_{1T}^{\perp g}(x,\bfp^{2}).
	\end{align}
	Using the above Eqs.~(\ref{eq:gtmd1})-(\ref{eq:gtmd3}), one can compute all the T-even TMDs. The unpolarized TMD, $f_{1}^{g}(x,\bfp^{2})$ is defined as the overlap representation of the proton light-front wave functions as \cite{More:2017zqp}
	\begin{eqnarray} \label{unpolTMDoverlap}\nonumber
		f_{1}^{g}(x,\bfp^{2})&=&\frac{1}{16\pi^{3}}\sum_{\lambda_{g}\lambda_{X}}(\epsilon^{1\ast}_{\lambda_{g}}\epsilon^{1}_{\lambda_{g}}+\epsilon^{2\ast}_{\lambda_{g}}\epsilon^{2}_{\lambda_{g}})\psi_{\lambda_{g}\lambda_{X}}^{\uparrow\ast}(x,\bfp^{2})\psi_{\lambda_{g}\lambda_{X}}^{\uparrow}(x,\bfp^{2})\\
		&=&\frac{1}{16\pi^{3}}\bigg[|\psi_{+1 +1/2}^{\uparrow}(x,\bfp^{2})|^{2}+|\psi_{+1 -1/2}^{\uparrow}(x,\bfp^{2})|^{2}+|\psi_{-1 +1/2}^{\uparrow}(x,\bfp^{2})|^{2} \bigg].
	\end{eqnarray}
	After employing the light-front wave functions,  Eqs.~(\ref{LFWFsuparrow}) and (\ref{LFWFsdownarrow}) in the above  Eq.~(\ref{unpolTMDoverlap}), we obtain the gluon unpolarized TMD as, 
	\begin{eqnarray}\label{unpolTMD}
		f_{1}^{g}(x,\bfp^{2})=N_{g}^{2}\frac{2}{\pi\kappa^2}\frac{\log[1/(1-x)]}{x}x^{2b}(1-x)^{2a}\bigg[ A(x)+\bfp^{2} B(x) \bigg]\exp[-C(x)\bfp^{2}],
	\end{eqnarray}
	where $A(x)$, $B(x)$ and $C(x)$ are given by 
	\begin{eqnarray}\label{UnpolTMT2}
		A(x)=\bigg(M-\frac{M_{X}}{(1-x)}\bigg)^{2}, \hspace*{0.5cm} B(x)=\frac{1+(1-x)^{2}}{x^{2}(1-x)^{2}}\hspace*{0.5cm} \text{and} \hspace*{0.5cm} C(x)=\frac{\log[1/(1-x)]}{\kappa^{2}x^{2}}.
	\end{eqnarray}
	Similarly, the gluon helicity TMD $g_{1L}^{g}(x,\bfp^{2})$, which describes the distribution of a circularly polarised gluon in a longitudinally polarised proton, is defined as 
	\begin{eqnarray}\label{helicityTMDoverlap}
		g_{1L}^{g}(x,\bfp^{2})&=&\frac{1}{16\pi^{3}}i\sum_{\lambda_{g}\lambda_{X}}(\epsilon^{2\ast}_{\lambda_{g}}\epsilon^{1}_{\lambda_{g}}-\epsilon^{1\ast}_{\lambda_{g}}\epsilon^{2}_{\lambda_{g}})\psi_{\lambda_{g}\lambda_{X}}^{\uparrow\ast}(x,\bfp^{2})\psi_{\lambda_{g}\lambda_{X}}^{\uparrow}(x,\bfp^{2})\\
		&=&\frac{1}{16\pi^{3}}\bigg[|\psi_{+1 +1/2}^{\uparrow}(x,\bfp^{2})|^{2}+|\psi_{+1 -1/2}^{\uparrow}(x,\bfp^{2})|^{2}-|\psi_{-1 +1/2}^{\uparrow}(x,\bfp^{2})|^{2} \bigg].
	\end{eqnarray}
	The analytical expression for the gluon helicity TMD in our model is obtained as,
	\begin{eqnarray}\label{helicityTMD}
		g_{1 L}^{g}(x,\bfp^{2})=N_{g}^{2}\frac{2}{\pi\kappa^2}\frac{\log[1/(1-x)]}{x}x^{2b}(1-x)^{2a}\bigg[ A(x)+\bfp^{2} \Tilde{B}(x) \bigg]\exp[-C(x)\bfp^{2}],
	\end{eqnarray}
	where $A(x)$, $C(x)$ are same as (\ref{UnpolTMT2}), while $\Tilde{B}(x)$ is given as 
	\begin{eqnarray}\label{Bprime}
		\Tilde{B}(x)&=&\frac{1-(1-x)^{2}}{x^{2}(1-x)^{2}}.
	\end{eqnarray}	
	The worm-gear gluon TMD $g_{1T}^{g}(x,\bfp^2)$ is defined as the distribution of a circularly polarised gluon in a transversely polarized proton~\cite{Lyubovitskij:2020xqj} and given by,
	\begin{eqnarray}\nonumber
		\frac{\bfp.\mathbf{S_{\perp}}}{M}g_{1T}^{g}(x,\bfp^2)&=& \frac{1}{16\pi^{3}}\frac{i}{2}\sum_{\lambda_{g}\lambda_{X}}(\epsilon^{1\ast}_{\lambda_{g}}\epsilon_{\lambda_{g}}^{2}-\epsilon^{2\ast}_{\lambda_{g}}\epsilon^{1}_{\lambda_{g}})\bigg[\psi_{\lambda_{g}\lambda_{X}}^{\uparrow\ast }(x,\bfp)\psi_{\lambda_{g}\lambda_{X}}^{\downarrow}(x,\bfp)+\psi_{\lambda_{g}\lambda_{X}}^{\downarrow \ast }(x,\bfp)\psi_{\lambda_{g}\lambda_{X}}^{\uparrow}(x,\bfp)\bigg]\\ 
		g_{1T}^{g}(x,\bfp^{2})&=& -\frac{4M}{16\pi^{3}x}\bigg(M-\frac{M_{X}}{(1-x)}\bigg)[\varphi(x,\bfp^{2})]^{2}.
	\end{eqnarray}
	 Using the soft-wall AdS/QCD wave function~(\ref{AdSphi}), the above equation can be written as,
	\begin{eqnarray}\label{wormgearTMD}
		g_{1T}^{g}(x,\bfp^{2})&=& -\frac{4M}{\pi \kappa^{2}}N_{g}^{2}\bigg(M(1-x)-{M_{X}}\bigg)\log[1/(1-x)]x^{2b-2}(1-x)^{2a-1}\exp[-C(x)\bfp^{2}].
	\end{eqnarray}
	Finally, The Boer-Mulders gluon TMD $h_{1}^{\perp g}(x,\bfp^{2})$, which describes a linearly polarized gluon inside an unpolarized proton, is given as,
	\begin{eqnarray}\nonumber\label{BMTMD}
		\frac{\bfp^{2}}{2M^{2}}h_{1}^{\perp g}(x,\bfp^2)&=& -\frac{1}{2}\sum_{\lambda_{N}\lambda_{g}\neq\lambda_{g}^{\prime}\lambda_{X}}\frac{1}{16\pi^{3}}\Big[\psi_{\lambda_{g}^{\prime}\lambda_{X}}^{ \star\lambda_{N}}(x,\bfp)\psi_{\lambda_{g}\lambda_{X}}^{\lambda_{N}}(x,\bfp)\epsilon^{\mu\dagger }_{\lambda_{g}^{\prime}}\epsilon^{\nu}_{\lambda_{g}}\Big],\, \\ \nonumber
		h_{1}^{\perp g}(x,\bfp^{2})&=&-\frac{1}{16\pi^{3}}\frac{M^{2}}{\bfp^{4}}\sum_{\lambda_{N}\lambda_{X}}\bigg[(p^{1}-ip^{2})^{2}\psi_{+1\lambda_{X}}^{\star\lambda_{N}}(x,\bfp)\psi_{-1\lambda_{X}}^{\lambda_{N}}(x,\bfp)+(p^{1}+ip^{2})^{2}\psi_{-1\lambda_{X}}^{\star\lambda_{N}}(x,\bfp)\psi_{+1\lambda_{X}}^{\lambda_{N}}(x,\bfp)\bigg]\\ \nonumber
		&=& \frac{8M^{2}}{16\pi^{3}}\frac{1}{x^{2}(1-x)}[\varphi(x,\bfp^{2})]^{2},\\ \nonumber
		h_{1}^{\perp g}(x,\bfp^{2})&=&\frac{8M^{2}}{\pi\kappa^{2}} N_{g}^{2}\log[1/(1-x)]x^{2b-3}(1-x)^{2a-1}\exp[-C(x)\bfp^{2}].
	\end{eqnarray}
   After performing the  $\bfp$-integration of the gluon unpolarized TMD, Eq.~(\ref{unpolTMD}), we obtain the corresponding collinear unpolarized PDF, $f_{1}^{g}(x)$ as, 
\begin{eqnarray}
	f_{1}^{g}(x)&=&\int d^{2}\bfp f_{1}^{g}(x,\bfp^{2}) \\ \nonumber
	&=&2N_{g}^{2} x^{2b+1}(1-x)^{2a-2}\bigg[ \kappa^2\frac{(1+(1-x)^2)}{\log[1/(1-x)]}+(M(1-x)-M_x)^2 \bigg].
\end{eqnarray} 
The gluon helicity PDF $g_{1L}^{g}(x)$ can be obtained after the $\bfp$-integration of the gluon helicity TMD in Eq.~(\ref{helicityTMD}) as,
\begin{eqnarray}
	g_{1L}^{g}(x)&=&\int d^{2}\bfp g_{1L}^{g}(x,\bfp^{2}) \\ \nonumber
		&=&2N_{g}^{2} x^{2b+1}(1-x)^{2a-2}\bigg[ \kappa^2\frac{(1-(1-x)^2)}{\log[1/(1-x)]}+(M(1-x)-M_x)^2 \bigg].
\end{eqnarray}
Similarly, the collinear PDFs of worm-gear, Eq.(\ref{wormgearTMD}), and the Boer-Mulders, Eq.(\ref{BMTMD}), TMDs are given as,
\begin{eqnarray}
	g_{1T}^{g}(x)&=&\int d^{2}\bfp g_{1T}^{g}(x,\bfp^{2}) \\ \nonumber
		&=&4M N_{g}^{2}\big(M(1-x)-M_{x}\big)x^{2b}(1-x)^{2a-1},
\end{eqnarray}
and,
\begin{eqnarray}
	h_{1}^{\perp g}(x)&=&\int d^{2}\bfp h_{1}^{\perp g}(x,\bfp^{2}) \\ \nonumber
	&=&8MN_{g}^{2}x^{2b-1}(1-x)^{2a-1}.
\end{eqnarray}
	\section{Numerical fitting and model parameters}\label{sec:fitting}
 There are four parameters $a$, $b$, $N_g$, and $M_X$ in our model, which will decide the goodness of the model. The parameters $N_g$ and $M_X$ are free parameters and they are fixed by normalization conditions of the gluon PDFs and spectator mass properties of the proton, respectively. The parameters $a$ and $b$ decide the behaviour of the distributions in extreme limits of $x $ are crucial to fix. We determine these model parameters, by fitting our unpolarized gluon distribution with the latest available gluon PDF data  at NLO of the gluon distribution $ x f_1^g(x)$ from the global analysis by the NNPDF Collaboration \cite{NNPDF:2017mvq}.  We particularly fit NNPDF3.0 NLO unpolarized gluon distribution at the scale $Q_0=2 $ GeV. We choose 300 data points within the interval $0.001 < x < 1 $ and 100 replicas of the gluon distribution. The effective uncertainties are calculated from the standard deviation of these 100 replicas for each value of $x_{i}$.

We set the gluon mass $M_g=0$.
 The choice of model parameters $a$ and $b$ depend on the spectator mass. During the search for the optimal fit, we find that for spectator mass close to the proton mass, the  model parameters produce a more physically acceptable spin contribution of the gluon than for larger spectator mass. Here, we choose $M_{X}=0.985$ GeV. In a similar kind of spectator model, the spectator mass has been chosen as $M_{x}=0.943$~\cite{Lu:2016vqu}. The model is very sensitive in the small $x$ region. Even in the NNPDF analysis, the polarised PDF has large uncertainty in the small-$x$ region, which makes the spin contribution prediction sensitive to the lower limit of $x$. Keeping all of this in mind, we exclude a very small $x$ region  from our fitting and our model is valid for the range $0.001<x<1$. 
	
\begin{figure}
		\includegraphics[scale=0.5]{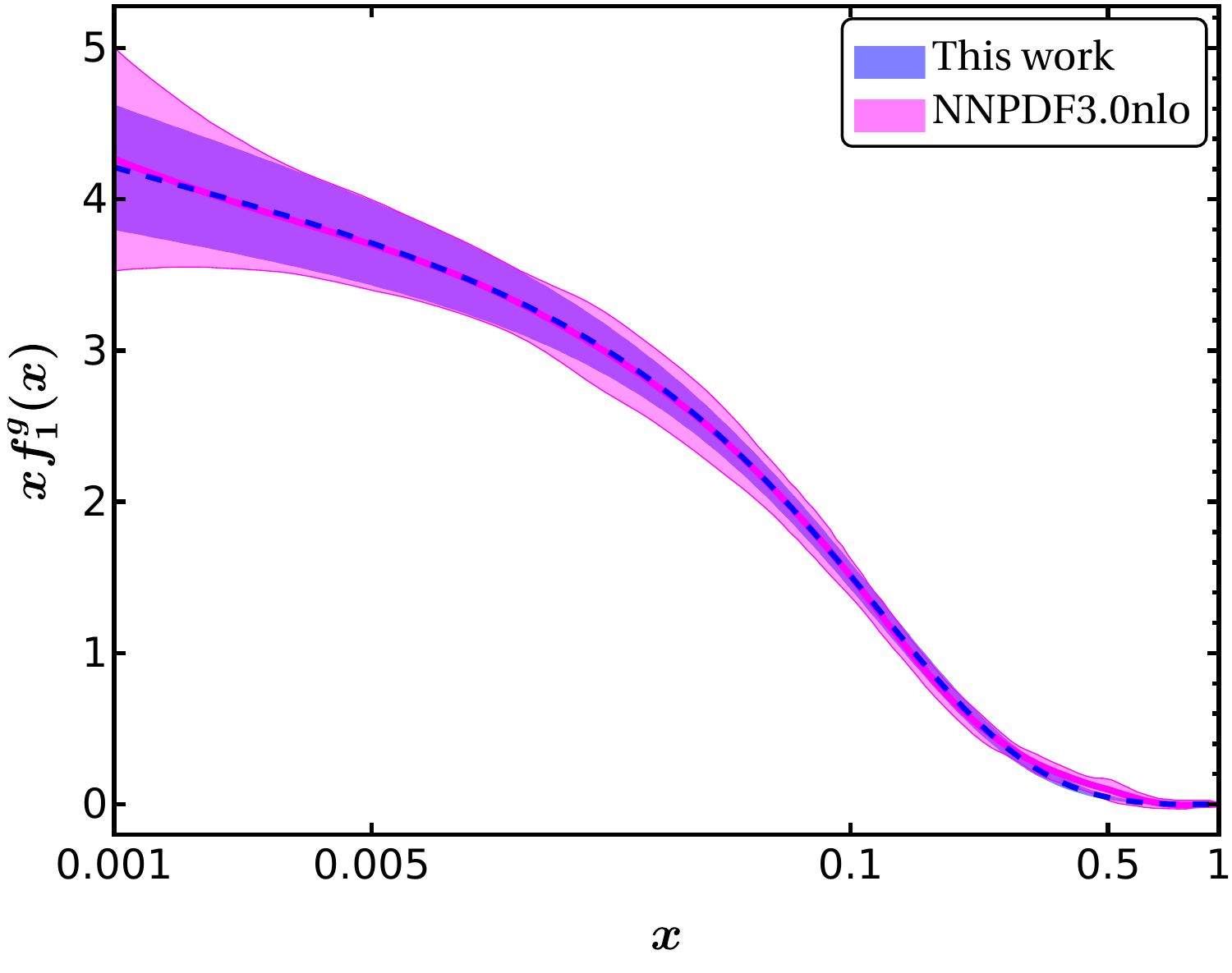}
		\caption{Our model Unpolarized gluon PDF $f_1^g(x)$ (blue dashed line with blue band of $2\sigma$ error) compared with the NNPDF3.0nlo data set (solid magenta line with magenta band) as a function of longitudinal momentum fraction $x$ in the kinematics region $0.001 \leq x \leq 1 $ at $Q_0=2$GeV.}
		\label{fig:gluonpdfs}
	\end{figure}

	The value of the fitted model parameters is listed in Table~\ref{Tab:modelparameters}.
		\begin{table}[b]
			\caption{Numerical values and the uncertainties of the fitted
				model parameters $a$ and $b$.}
			\label{Tab:modelparameters}
			\centering
		\begin{tabular}{ |c|c|c|c|c| } 
			\hline\hline
			Parameter & Central Value &  $ 1\sigma $-Error band&  $2 \sigma $-Error band \\
			\hline
			$a$	& 3.88&  $\pm$ 0.1020 & $\pm$ 0.2232 \\ 
			$b$	& -0.53 &  $\pm $ 0.0035&  $\pm $ 0.0071 \\ 
				\hline \hline
		\end{tabular}
	\end{table}
These model parameters are fixed by fitting the NNPDF3.0 NLO data set at $\mu^{2}=4$ GeV$^2$ with a $\chi_{\rm min}^{2}=20.37$ with the normalization constant $N_g = 2.088$. 
 We notice that the $2 \sigma$ uncertainty to the parameter fitting is close to the experimental error corridor, and we take $2 \sigma$ uncertainty as a standard maximized error in this model for further reporting.  The parameters in the wave functions determined by the fitting of unpolarized gluon PDF can be further employed  to predict the other gluon distributions e.g., gluon helicity, transversity, TMDs etc. In Fig.~\ref{fig:gluonpdfs}, we show the results of our fit for the unpolarized gluon distribution $xf_{1}^{g}(x)$ at $Q_{0} = 2$ GeV. The solid magenta  band identifies	the NNPDF3.0 parametrization of $xf_{1}^{g}(x)$~\cite{Ball:2017otu} and the blue-dashed line with the blue band shows our model results at $2 \sigma$ error corridor. 
	
\section{Results}\label{sec:Results}

The value of the average longitudinal momentum of the gluon is defined as the second Mellin's moment of the unpolarized PDF as,
\begin{eqnarray}
\langle x\rangle_{g}=\int_{0}^{1}dx x f_{1}^{g}(x)=0.416^{+0.048}_{-0.041},
\end{eqnarray} 
which is close to the 
 recent lattice calculations at $Q^2=4$ GeV$^2$, $	\langle x\rangle_{g} = 0.427(92)$~\cite{Alexandrou:2020sml}. In Table~\ref{Tab:avgx}, we compared the average value of the longitudinal momentum fraction for the unpolarized gluon PDF with the available theoretical models in the literature ~\cite{Bacchetta:2020vty,Lyubovitskij:2020xqj,Lu:2016vqu,Kaur:2019kpe}.
\begin{table}
		\caption{Comparison of the numerical values of the average longitudinal momentum of the gluon at $Q_{0}=2 $ GeV.}
		\label{Tab:spincontributions}
		\centering
		\begin{tabular}{ |c|c|c|c|c|c|c|} 
			\hline\hline
	& This work & \cite{Bacchetta:2020vty} & \cite{Lu:2016vqu} & 
     \cite{Kaur:2019kpe} &
     \cite{Alexandrou:2020sml}  \\
			\hline
			$\langle x\rangle_{g}$ & 0.416 & 0.424 & 0.411& 0.409 & 0.427\\
			\hline \hline
		\end{tabular}
\label{Tab:avgx}
	\end{table}
\begin{figure}
\includegraphics[scale=0.5]{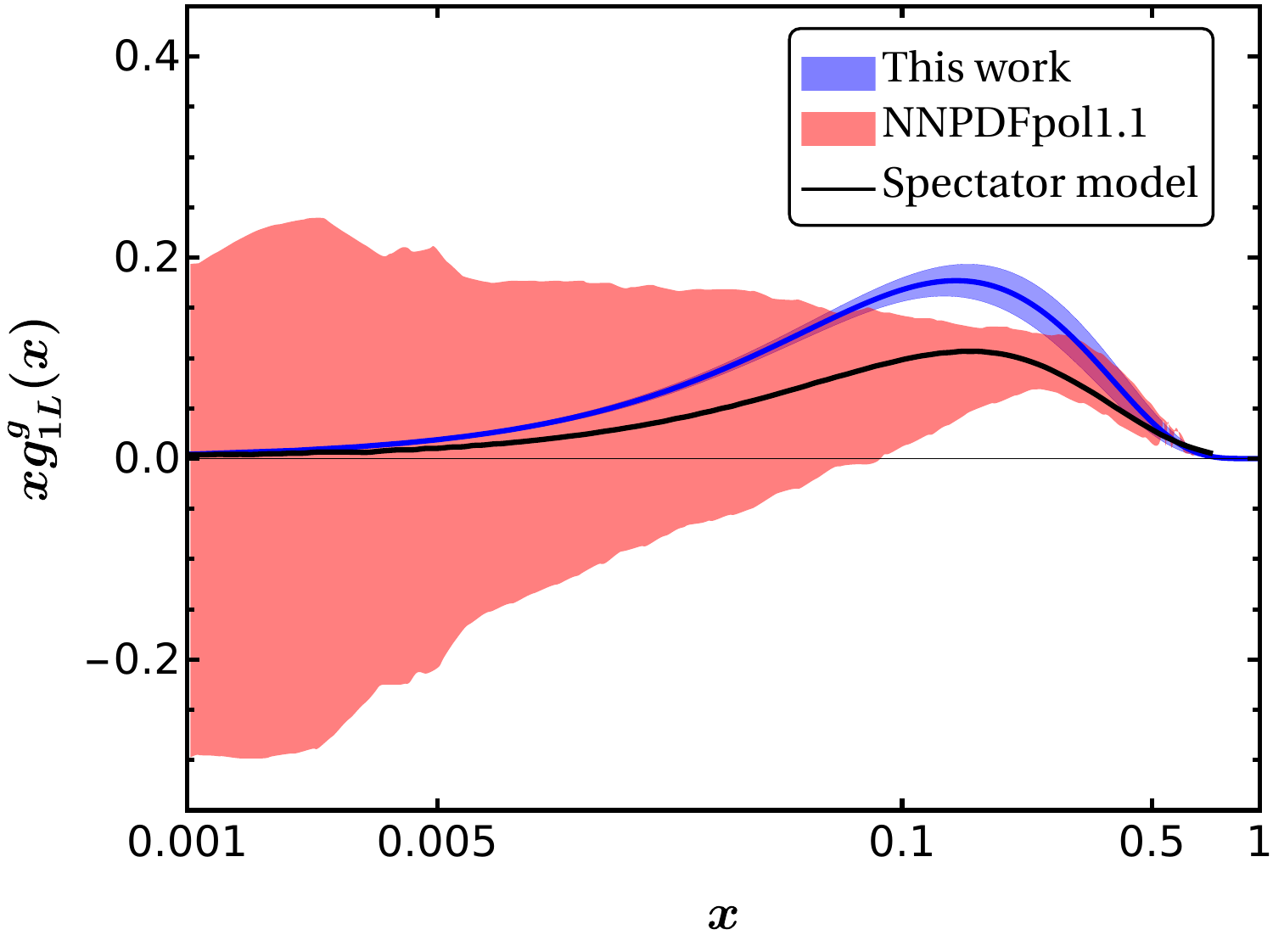}
	\hspace{0.5cm}
\includegraphics[scale=0.5]{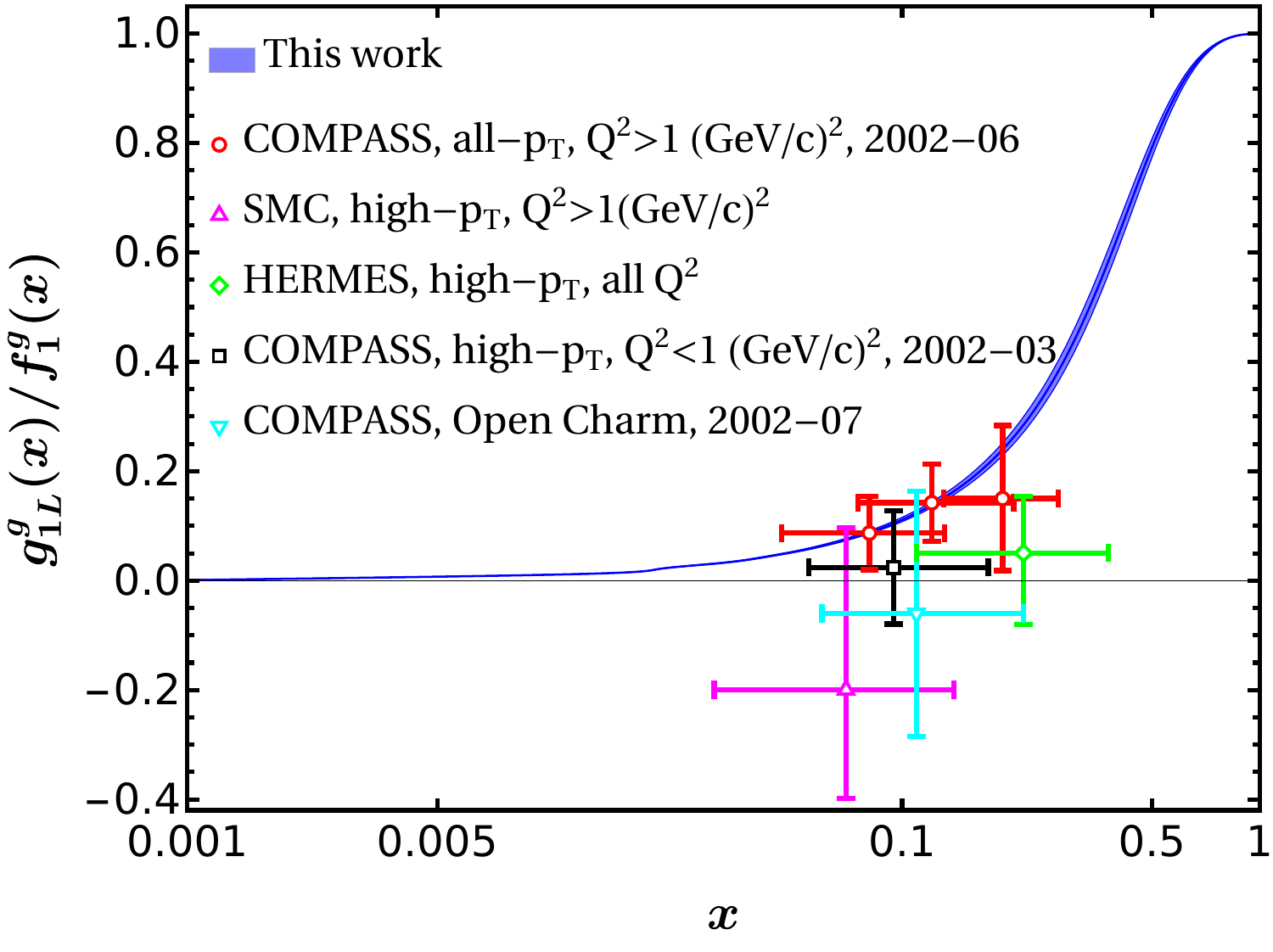}\hspace{0.5cm}
	\caption{The gluon helicity pdf, $g_{1L}^{g}(x)$ (left panel) compared with the NNPDFpol1.1~\cite{Nocera:2014gqa} and the spectator model results~\cite{Bacchetta:2020vty} at $Q_{0} = 2$ GeV. The right panel shows the comparison for the helicity asymmetry $g_{1L}^{g}(x)/f_{1}^{g}(x)$ from our calculation (blue band) with the experimental measurements. The direct measurements of COMPASS~\cite{COMPASS:2005qpp,COMPASS:2015pim},
HERMES~\cite{HERMES:2010nas} and SMC~\cite{SpinMuonSMC:2004jrx} are obtained in the leading order from high $p_{T}$ hadrons while open charm muon production at COMPASS ~\cite{COMPASS:2012mpe} are taken from next-to-leading order at different values of $x$.}
	\label{fig:helicityasymmetry}
\end{figure}
	
 In Fig.~\ref{fig:helicityasymmetry}, we show our model predictions for the polarized gluon 
distribution $xg_{1L}^{g}(x)$ (left panel) and the gluon helicity asymmetry ratio  $g^g_{1L}(x)/f_{1}^{g}(x)$ (right panel) at  $Q_{0} = 2$ GeV. The red band in the left panel of Fig.~\ref{fig:helicityasymmetry} represents the NNPDFpol1.1 results, which have large uncertainty in the entire range of $x$ and particularly in the small-$x$ region. The central line of NNPDFpol1.1 data is negative in the $x$ close to $0.001$ region, while in our model, the gluon helicity distribution is always positive. Overall, we find that our gluon helicity distribution in the entire region of $x$ except the domain $0.07<x<0.3$ is more or less consistent with the global analysis. Within the domain $0.07<x<0.3$, the distribution is going beyond the uncertainty band. As a result, we  obtain the high value of gluon  spin contribution in the small-$x$ region as shown in Table~\ref{Tab:spincontributions}. 
 The spin contribution for large-$x$ mainly comes from the quark sector and we can not expect much contribution from gluons. In Table~\ref{Tab:spincontributions}, we list the dependence of the gluon helicity on the $x$ range and also  compare them  with the other model predictions of gluon helicity in  certain ranges of $x$. We observe that the maximum contribution to the gluon helicity comes from the small-$x$ region. 
 Compared to other model results, the gluon helicity contributions for different $x$ regions are found to be relatively larger in our model.
   The high gluon spin contribution has been reported in  Refs.~\cite{Joo:2019bzr,Kaur:2019kpe}. Meanwhile, in Ref.~\cite{Bacchetta:2020vty}, the gluon spin contribution is  relatively small, $s_{g} = 0.159 \pm 0.011$, which may be due to the fact that the unpolarized as well as helicity PDFs have been simultaneously fitted in that model. The latest lattice result of the gluon total angular momentum is reported to be $J_{g} = 0.187(46)$ at the scale $2 $ GeV~\cite{Alexandrou:2020sml}.
	
	\begin{table}
		\caption{Comparison of the numerical values of the gluon spin contribution with the available data at $Q_{0} = 2$ GeV.}
		\label{Tab:spincontributions}
		\centering
		\begin{tabular}{ |c|c|c|} 
			\hline\hline
			Gluon helicity & Central Value &  our predictions\\
			\hline
			$\Delta G=\int_{0.05}^{0.3}dx\Delta g(x)$	& 0.20~~~~\cite{PHENIX:2008swq}   & $0.28_{-0.037}^{+0.047}$ \\ 
			
			$\Delta$G=$\int_{0.05}^{0.2}dx\Delta g(x)$	& 0.23(6)~\cite{Nocera:2014gqa}& $0.22^{+0.033}_{-0.024}$ \\ 
			
			$\Delta$G=$\int_{0.05}^{1}dx\Delta g(x)$	& 0.19(6)~\cite{deFlorian:2014yva}& $0.326^{+0.066}_{-0.050}$ \\
			\hline\hline
		\end{tabular}
	\end{table}
 In the right panel of Fig.~\ref{fig:helicityasymmetry}, the gluon helicity asymmetry has been compared with available experimental data. From this comparison, we notice that our result for helicity asymmetry  is in 
 good agreement with the experimental measurements. In our model, the helicity asymmetry ratio is independent of the model parameters $a$ and $b$ and depends only on the spectator mass $M_{x}$, which satisfies the following model-independent QCD constraints~\cite{Brodsky:1989db, Brodsky:1994kg} ,
\begin{eqnarray}
\lim_{x\rightarrow 0}	\frac{g_{1L}^{g}(x)}{f_{1}^{g}(x)}=0, \hspace{0.5cm} {\rm and} \hspace{0.5cm} \lim_{x\rightarrow 1}	\frac{g_{1L}^{g}(x)}{f_{1}^{g}(x)}=1 .
\end{eqnarray}

\begin{figure}
	\centering
	\includegraphics[scale=0.5]{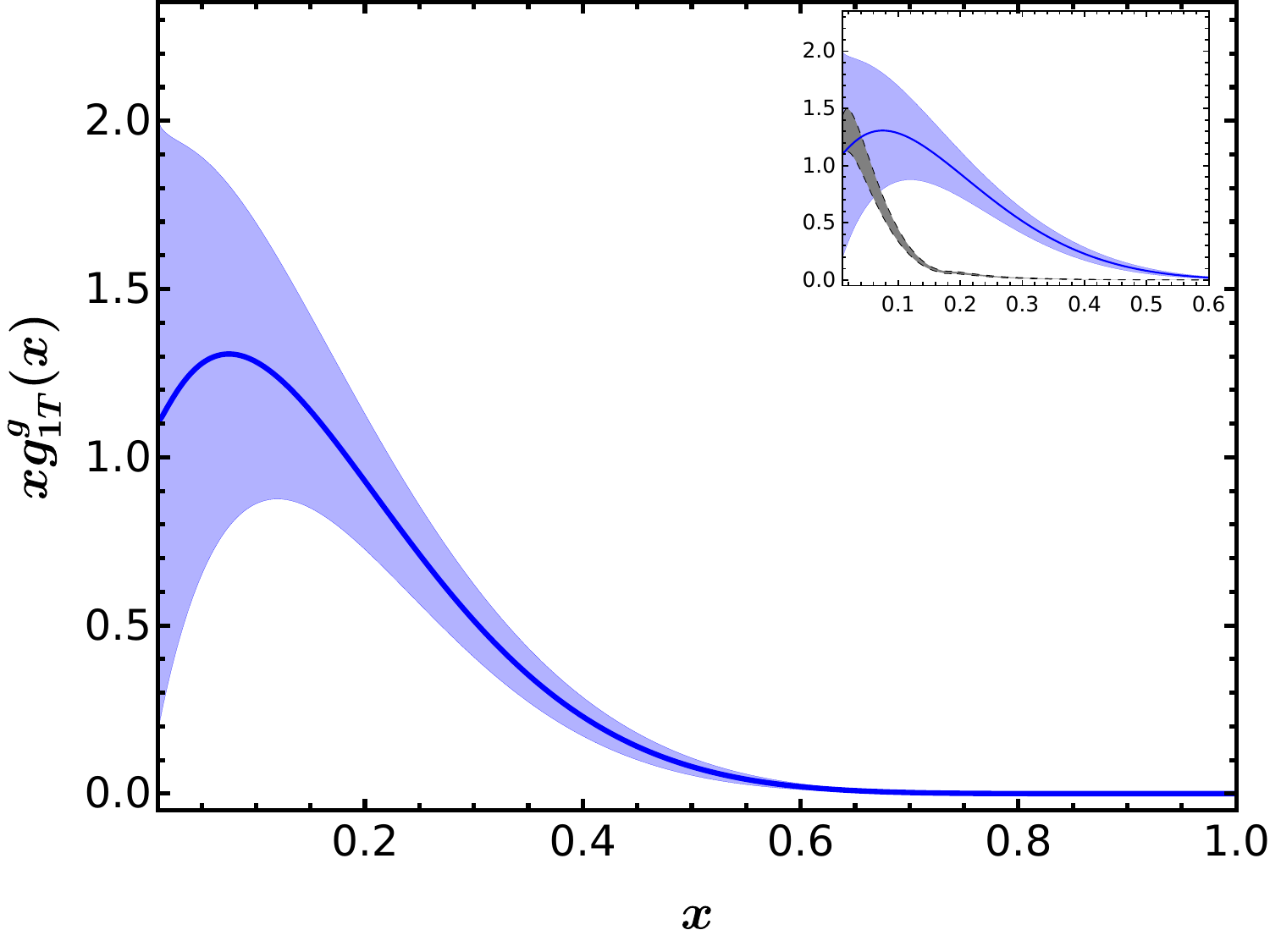}\hspace*{0.5cm}
	\includegraphics[scale=0.5]{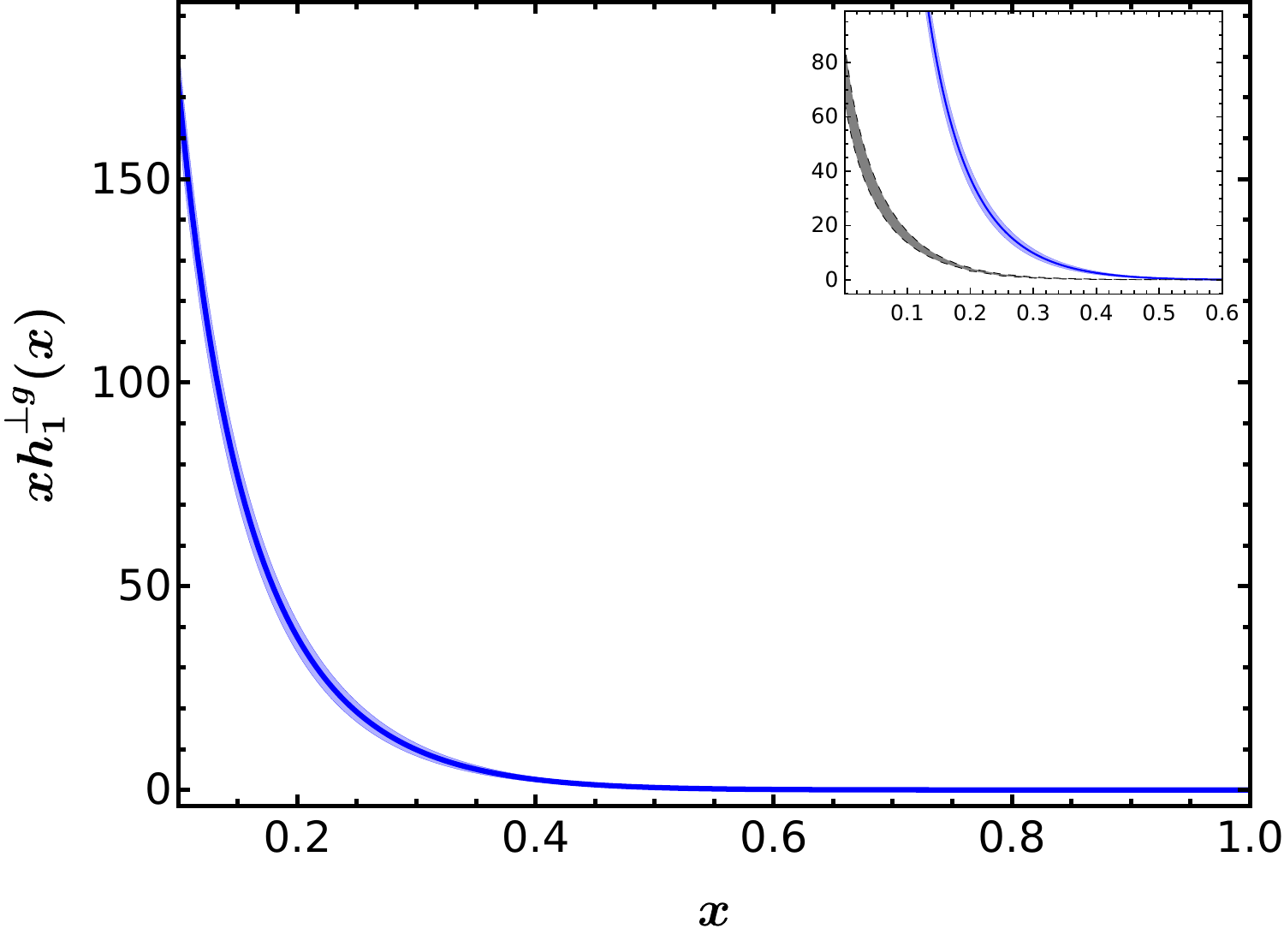}
	\caption{The gluon worm-gear, $xg_{1L}^{g}(x)$ (left) and the Boer-Mulders, $xh_{1}^{\perp g}(x)$ (right) collinear PDFs. The insets show a comparison of our results to those presented in Ref.~\cite{Lyubovitskij:2020xqj} with dashed borders.}
	\label{fig:Tevenpdfs}
\end{figure}
\begin{figure}
	\centering
	\includegraphics[scale=0.45]{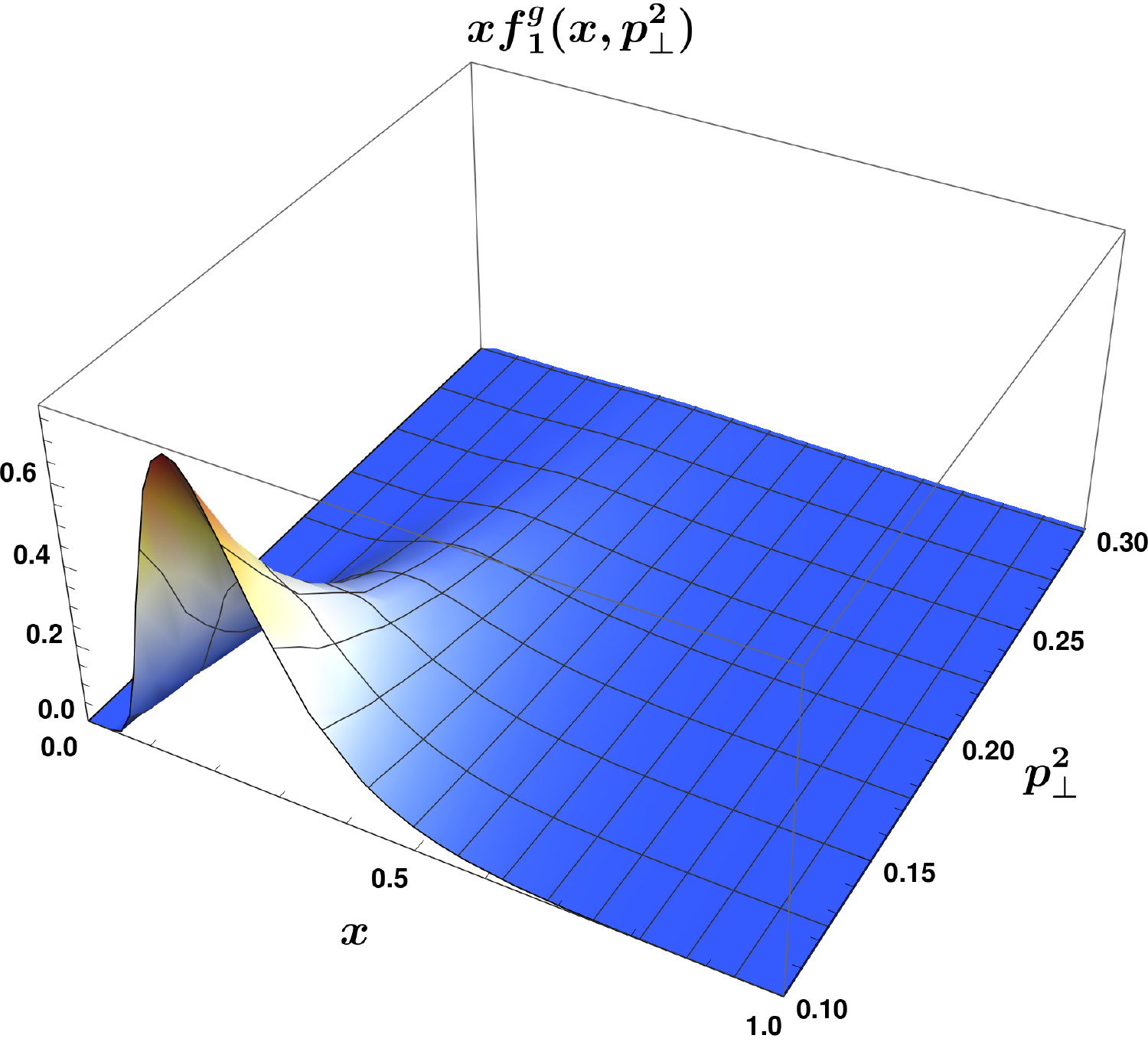}\hspace{0.5cm}
	\includegraphics[scale=0.45]{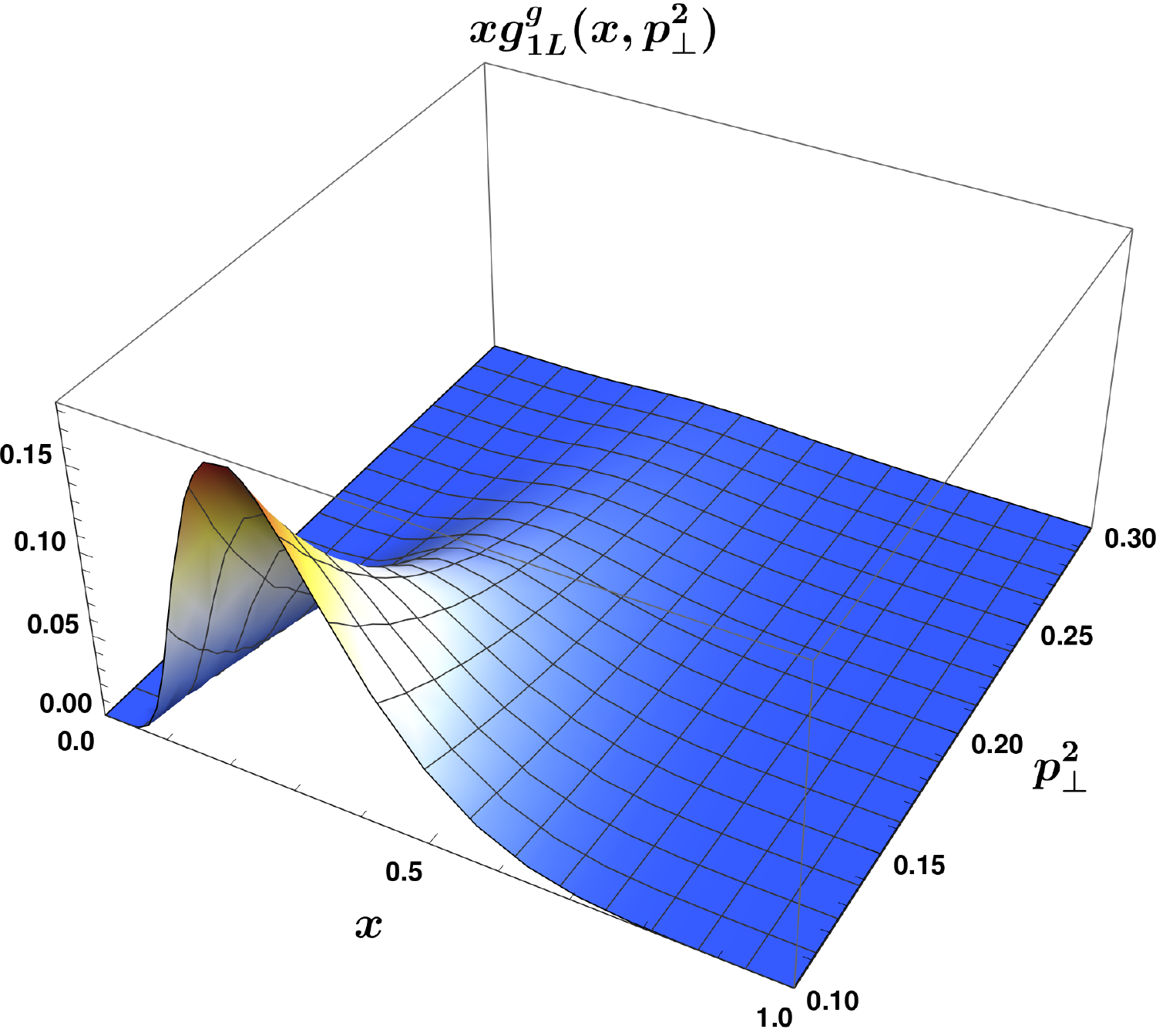}
	\vspace{0.5cm}
	\includegraphics[scale=0.45]{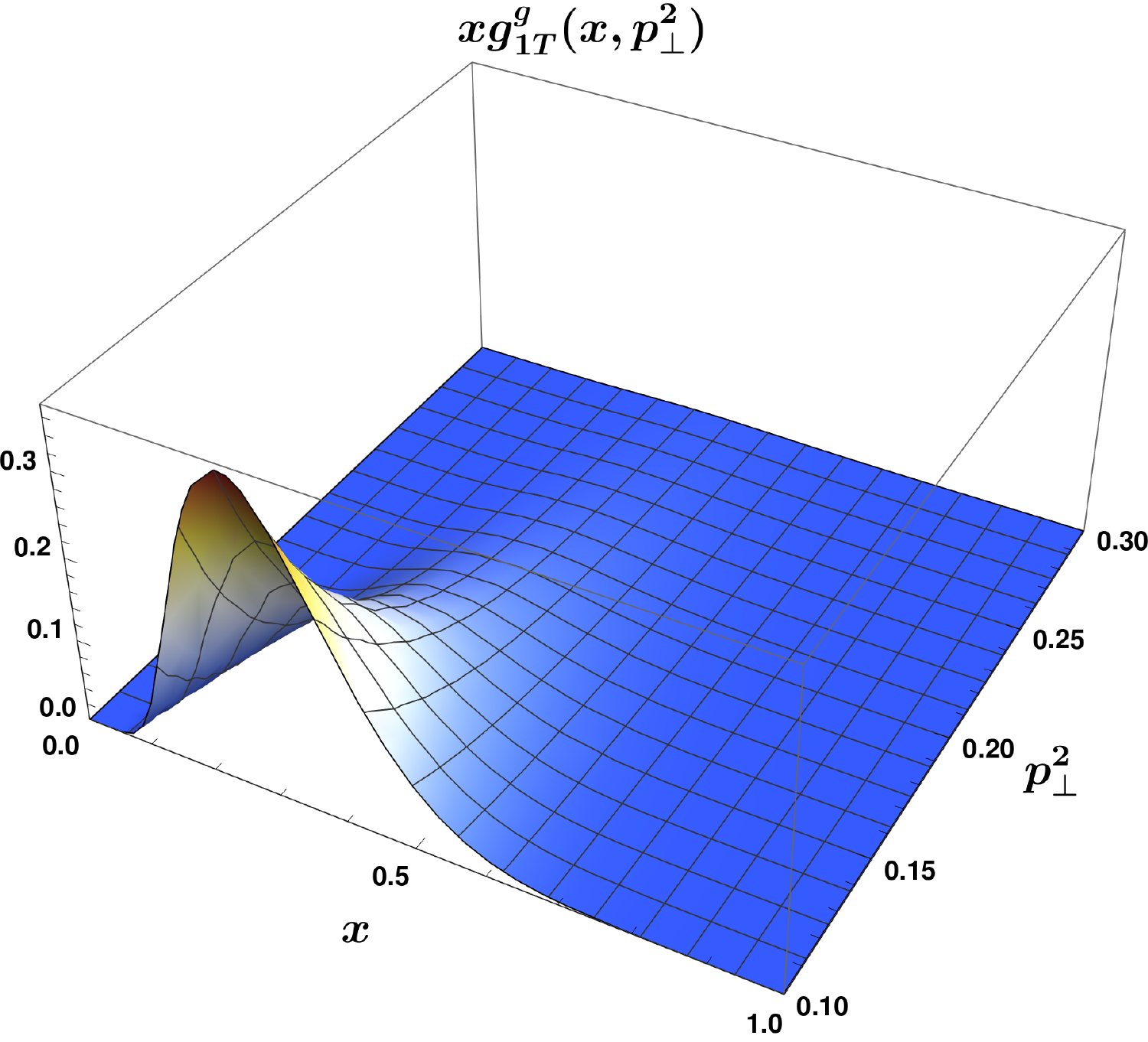}\hspace{0.5cm}
	\includegraphics[scale=0.45]{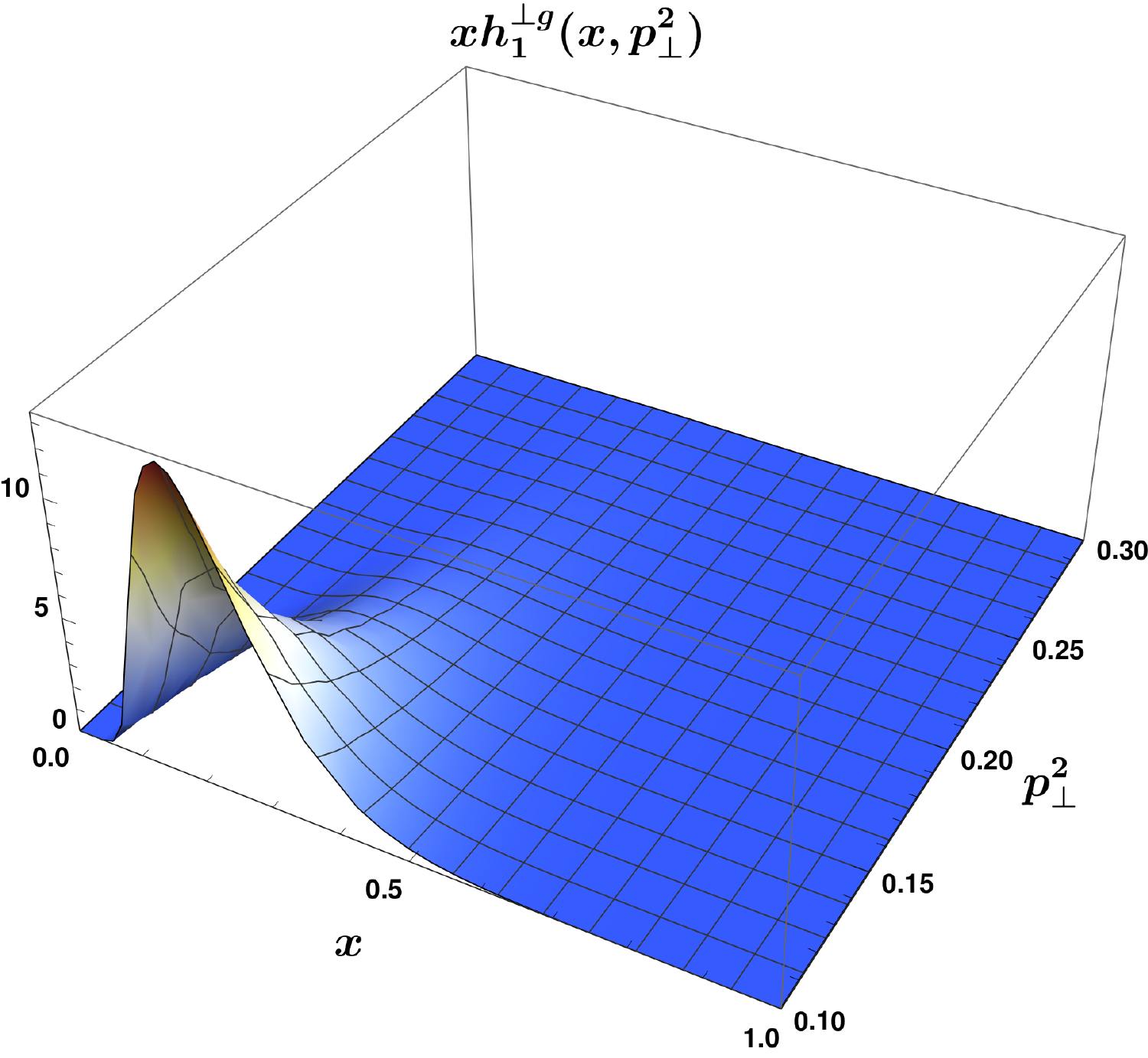}
	\caption{The TMDs for the gluon as functions of $x$ and $\bfp^{2}$. Upper panel: The unpolarized gluon TMD, $f_{1}^{g}(x,\bfp^{2})$ (left), The gluon helicity TMD, $g_{1L}^{g}(x,\bfp^{2})$ (right panel). Lower panel: The worm-gear TMD, $g_{1T}^{g}(x,\bfp^{2})$ (left) and the Boer-Mulders TMD, $h_{1}^{\perp g}(x,\bfp^{2})$ (right).}
	\label{fig:3DTMDs}
\end{figure}
The uncertainty band in the helicity asymmetry ratio plot (Fig.~\ref{fig:helicityasymmetry}) is created by including the errors in the spectator mass ($M_{x}=0.985^{+0.044}_{-0.045}$) in such a way that the maximum spin contribution should not go beyond the total proton spin and the lower cut-off for the spectator mass is $M_{x}\to M$.

In Fig.~\ref{fig:Tevenpdfs}, we show $x$-weighted collinear PDFs of worm-gear $g_{1T}^{g}(x)$ and the Boer-Mulders $h_{1}^{\perp g}(x)$ TMDs as a function of $x$. There is no  PDF corresponding to the  collinear limit of the worm-gear and the Boer-Mulders TMDs. We have also shown their comparison with the results reported in Ref.~\cite{Lyubovitskij:2020xqj} in the range $ 0< x< 0.6 $. 

\begin{figure}
	\centering
	\includegraphics[scale=0.5]{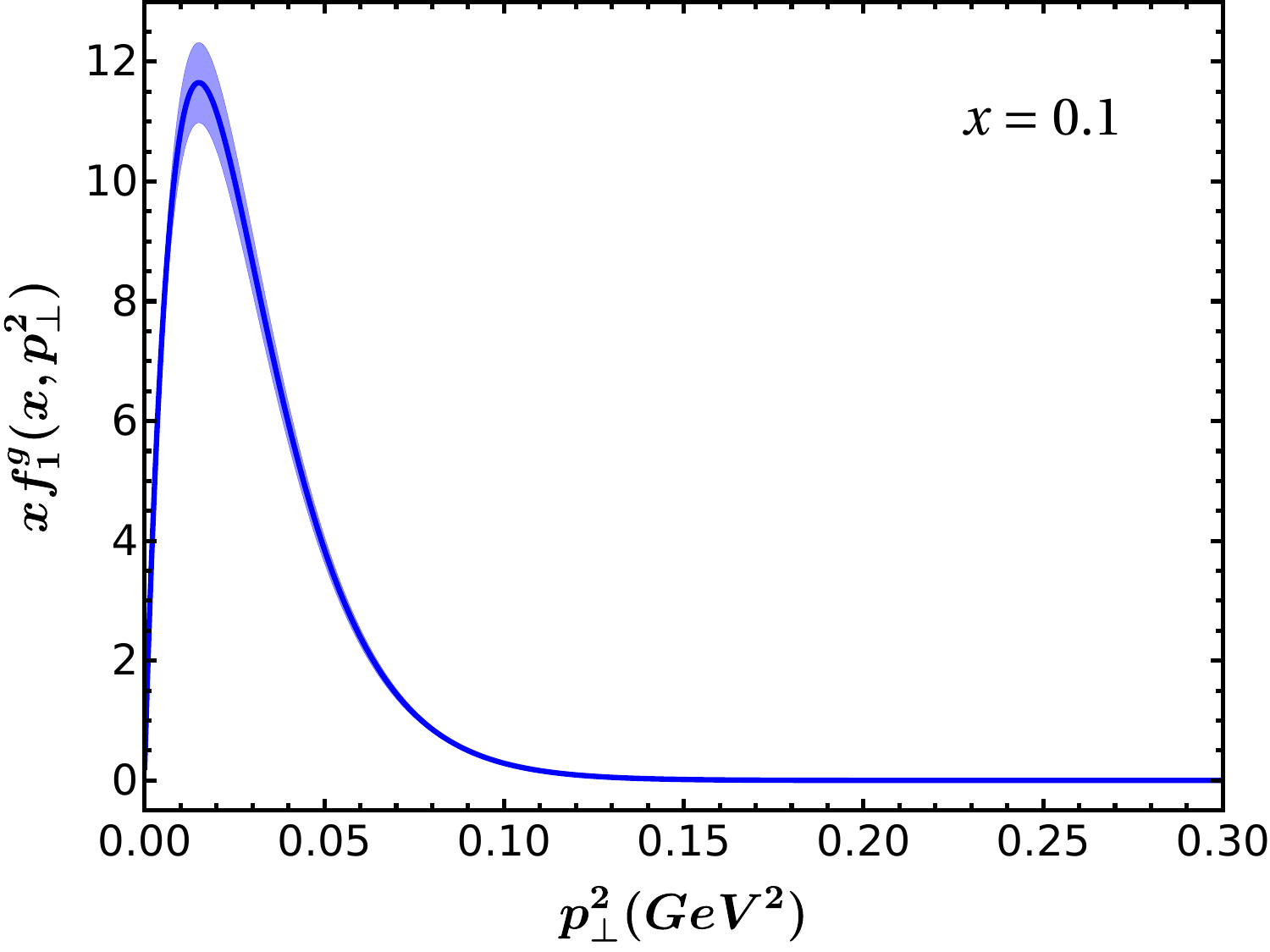}\hspace{0.5cm}
	\includegraphics[scale=0.5]{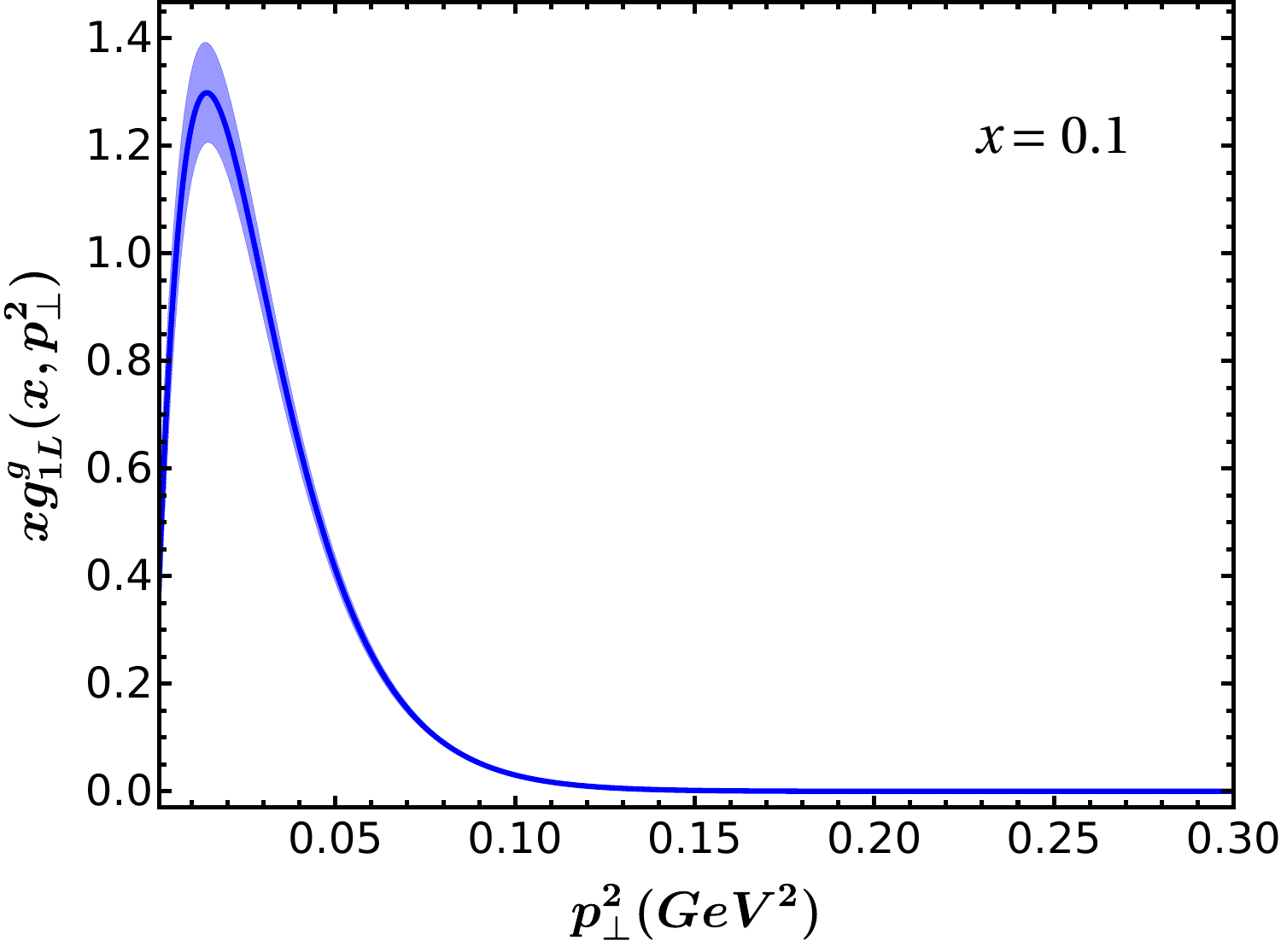}\vspace{0.5cm}
	\includegraphics[scale=0.5]{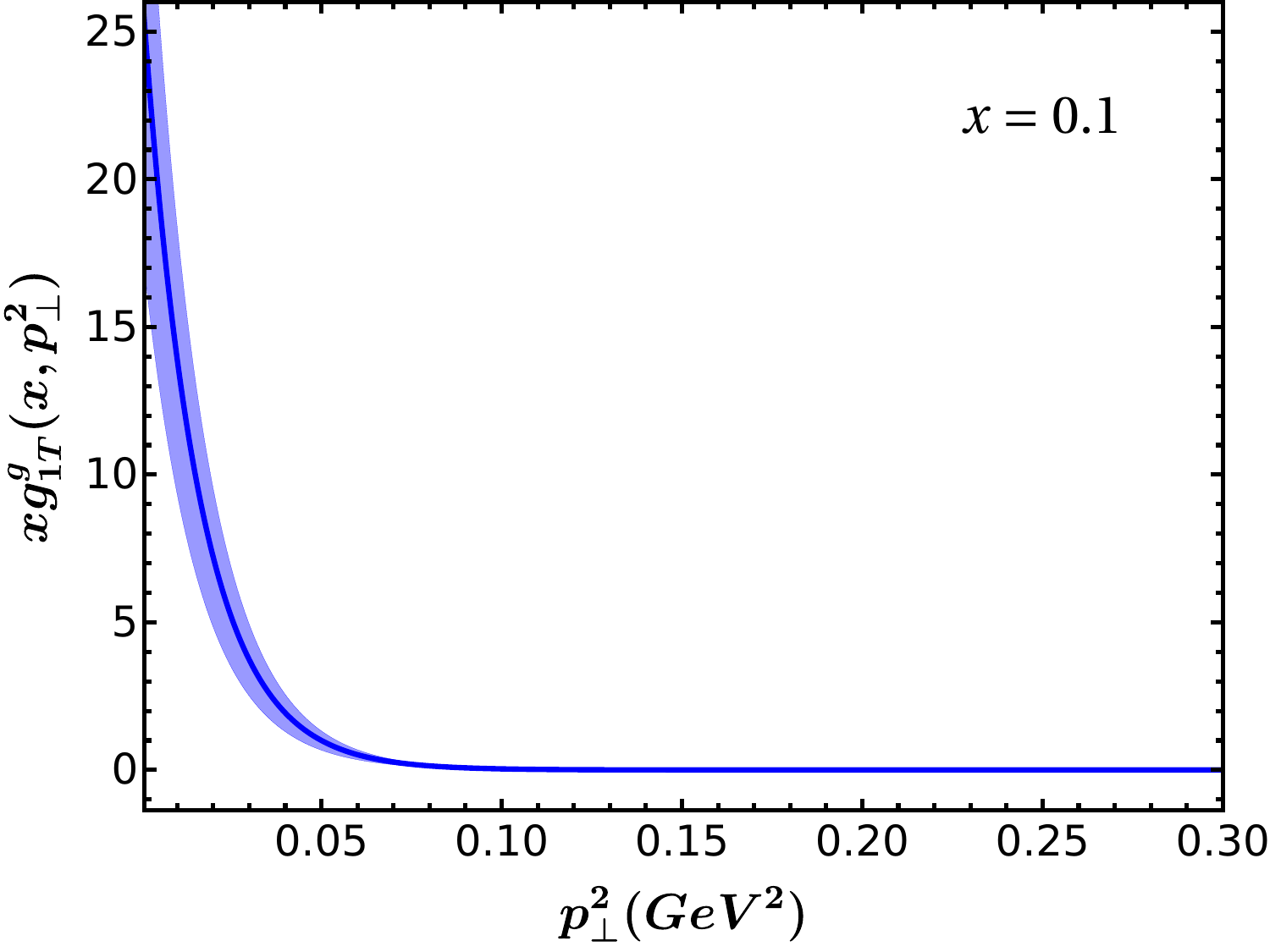}\hspace{0.5cm}
	\includegraphics[scale=0.5]{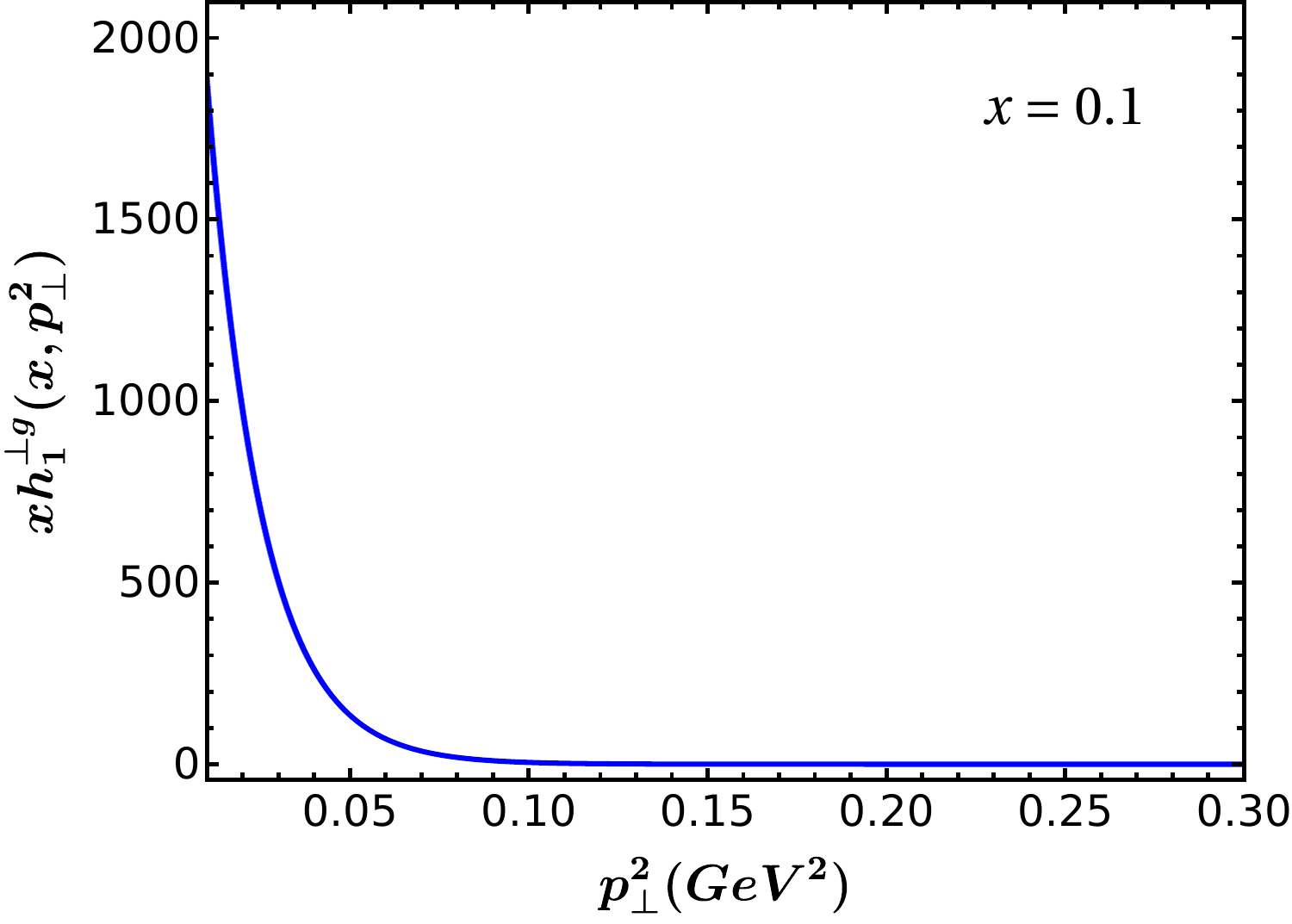}
	\caption{The T-even TMDs as a function of $\bfp^{2}$ for $x=0.1$. Upper panel: The unpolarized gluon TMD, $xf_{1}^{g}(x,\bfp^2)$ (left), The helicity TMD, $xg_{1L}^{g}(x,\bfp^{2})$ (right). Lower panel: The worm-gear TMD, $g_{1T}^{g}(x,\bfp^{2})$ (left) and the Boer-Mulders TMD, $h_{1}^{\perp g}(x,\bfp^{2})$ (right), respectively. }
	\label{fig:TevenTMDs}
\end{figure}

The three-dimensional distribution of the T-even TMDs, $f_{1}^{g}(x,\bfp^{2})$, $g_{1L}^{g}(x,\bfp^{2})$, $g_{1T}^{g}(x,\bfp^{2})$, and $h_{1}^{\perp g}(x,\bfp^{2})$  at the scale $Q_{0} = 2$ GeV are shown in Fig.~\ref{fig:3DTMDs}. All the T-even TMDs have their positive peaks around small $x$ and fall off very sharply with increasing $\bfp$. In Fig.~\ref{fig:TevenTMDs}, we present our model results for the T-even gluon TMDs as a function of $\bfp^{2}$ at $x=0.1$. These distributions are found to be slightly overestimated as compared to the results reported in Ref.~\cite{Lyubovitskij:2020xqj}, whereas the worm-gear TMD in Ref.~\cite{Bacchetta:2020vty} is shown to be negative.
We also notice that our model results for T-even TMDs fall off very sharply with $\bfp^{2}$ as compared to the other theoretical predictions~\cite{Lyubovitskij:2020xqj,Bacchetta:2020vty,Kaur:2019kpe}.  

  The following gluon densities are also pertinent since they describe the two-dimensional $\bfp$-distributions of gluons at various $x$ for various combinations of their polarization and nucleon spin state.
The unpolarized gluon density in an unpolarized nucleon is calculated as follows: 
 \begin{eqnarray}\label{eq:unpoldensity}
 	x\rho_g(x,p_{x},p_{y})=xf_{1}^{g}(x,\bfp^{2}),
 \end{eqnarray}
which describes the probability density of finding the unpolarized gluons 
at given $x$ and $\bfp$. The ``Boer-Mulders'' density, which shows the probability density of finding the linearly polarized gluons with $x$ and $\bfp$ is given as,
\begin{eqnarray}\label{eq:BMdensity}
	x\rho^{\leftrightarrow}_g(x,p_{x},p_{y})=\frac{1}{2}\bigg[xf_{1}^{g}(x,\bfp^{2})+\frac{p_{x}^{2}-p_{y}^{2}}{2M^{2}}xh_{1}^{\perp g}(x,\bfp^{2})\bigg].
\end{eqnarray}
Similarly, the ``helicity density'', which describes the probability density of circularly polarized gluons at particular $x$ and $\bfp$ inside the longitudinally polarized proton is given as,
\begin{eqnarray}\label{eq:helicitydensity}
	x\rho^{\circlearrowleft/+}_g(x,p_{x},p_{y})=xf_{1}^{g}(x,\bfp^{2})+xg_{1L}^{g}(x,\bfp^{2}).
\end{eqnarray}
Finally, the ``worm-gear density'', which describes the probability density of circularly polarized gluons at given $x$ and $\bfp$ inside the transversely polarized proton is given as,
\begin{eqnarray}
	x\rho^{\circlearrowleft/\leftrightarrow}_g(x,p_{x},p_{y})=xf_{1}^{g}(x,\bfp^{2})-\frac{p_{x}}{M}xg_{1T}^{g}(x,\bfp^{2}).
\end{eqnarray} 
The unpolarized, Eq.~(\ref{eq:unpoldensity}), and the helicity, Eq.~(\ref{eq:helicitydensity}), densities show that the $\bfp$ distributions are cylindrically symmetric around the longitudinal direction, 
as the proton (gluon) is unpolarized or longitudinally (circularly) polarized along to $P^{+}$.  The Boer-Mulders density Eq.~(\ref{eq:BMdensity}) is symmetric about the $p_{x}$ and $p_{y}$ axes because it describes unpolarized proton and linearly polarized gluons along the $p_{x}$ direction. The worm-gear density involves a transversely polarized proton along the $+p_{x}$ axes. Hence it is asymmetrically distributed in the same direction.
\begin{figure}
	\centering
	\includegraphics[scale=0.47]{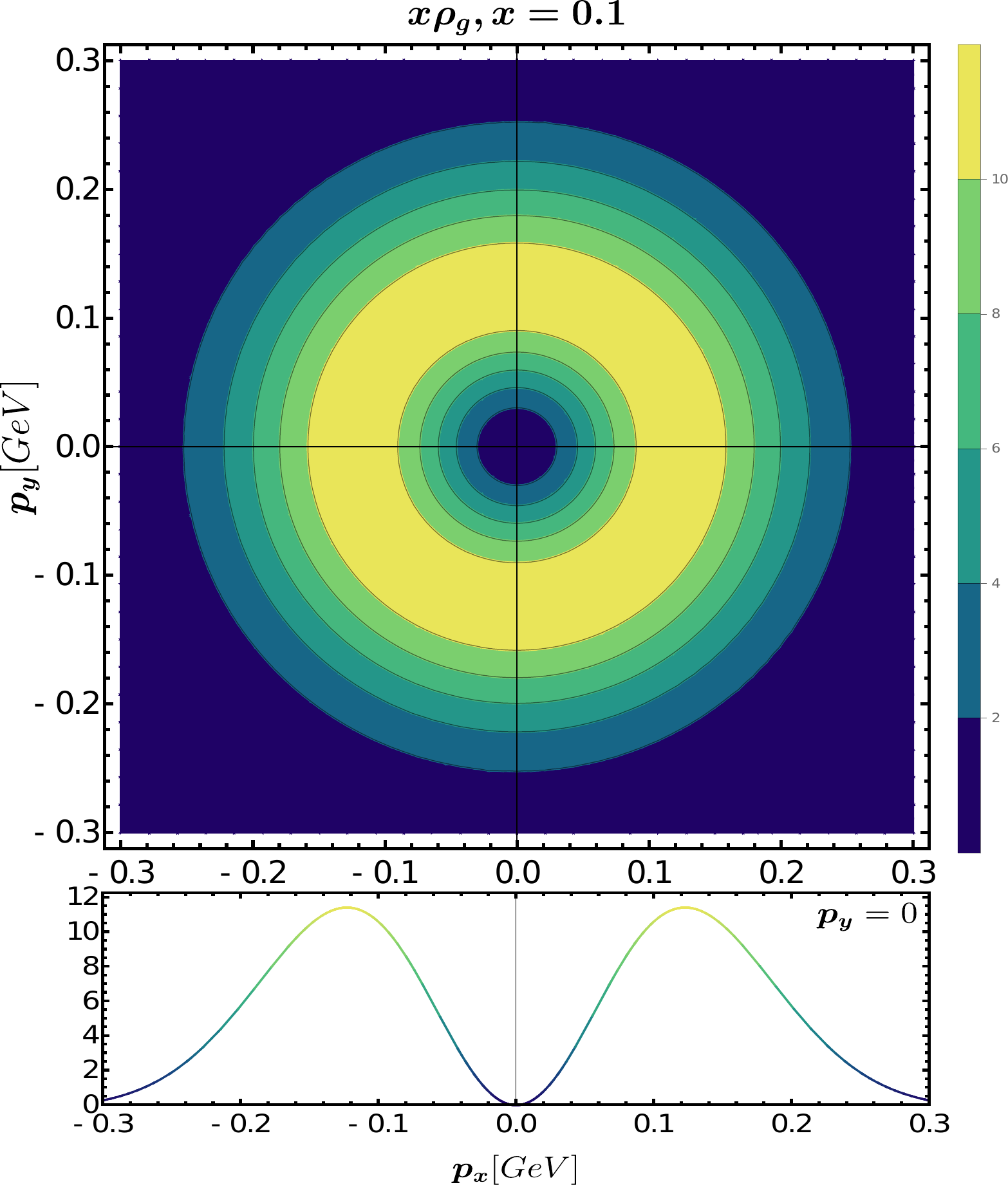}\hspace{0.5cm}
	\includegraphics[scale=0.47]{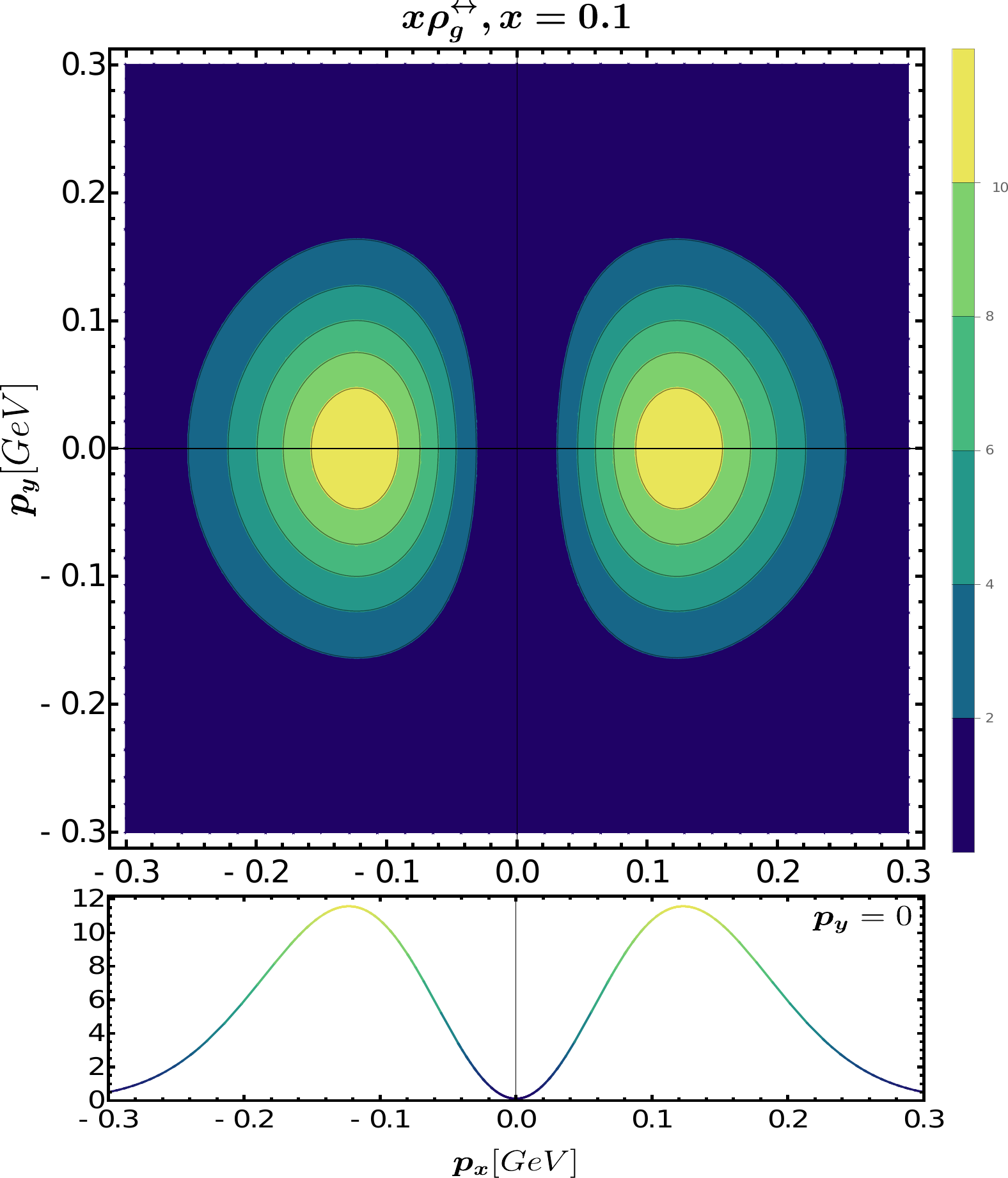}\\
	\vspace{0.4cm}
	\includegraphics[scale=0.47]{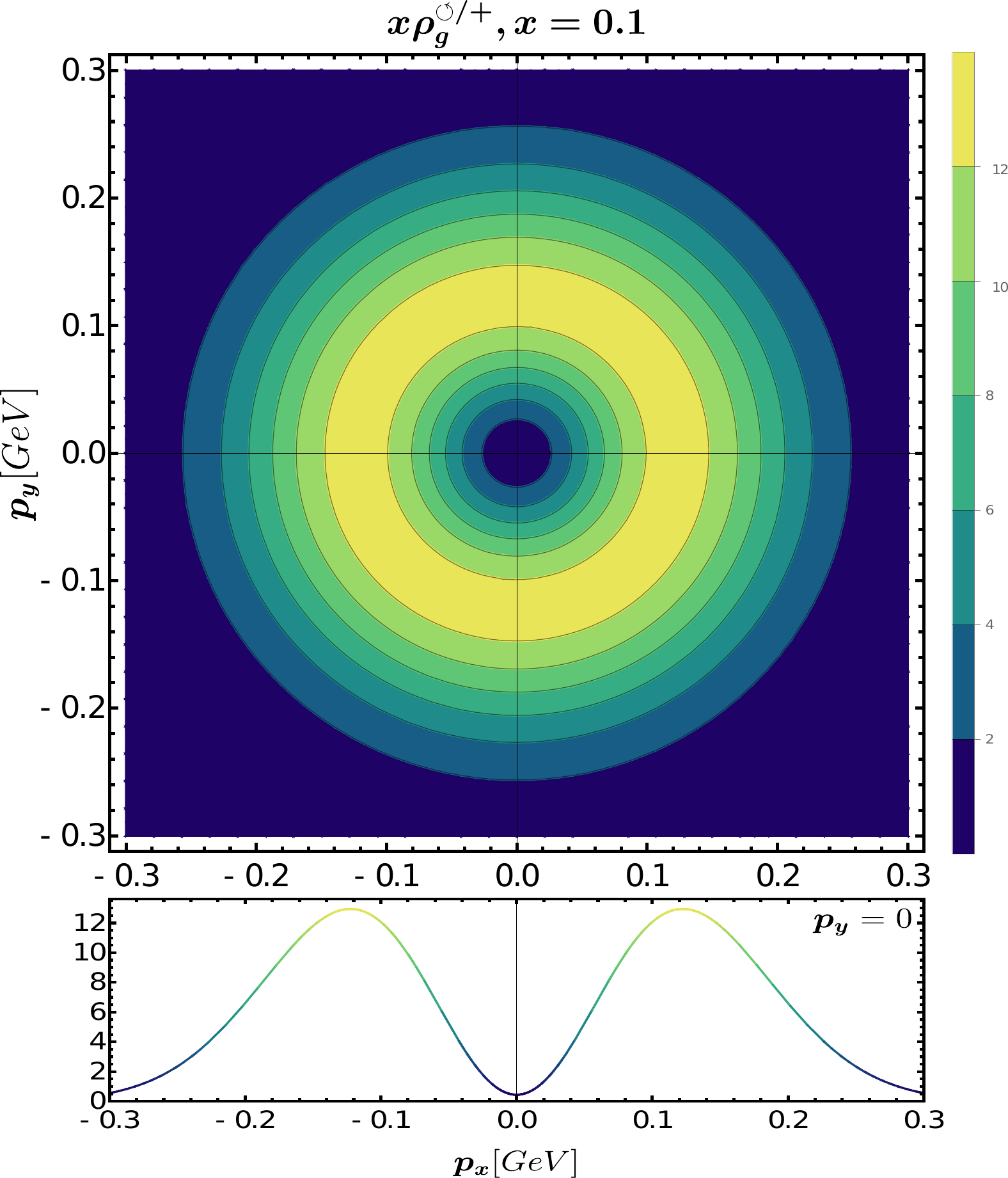}
	\hspace{0.5cm}
	\includegraphics[scale=0.47]{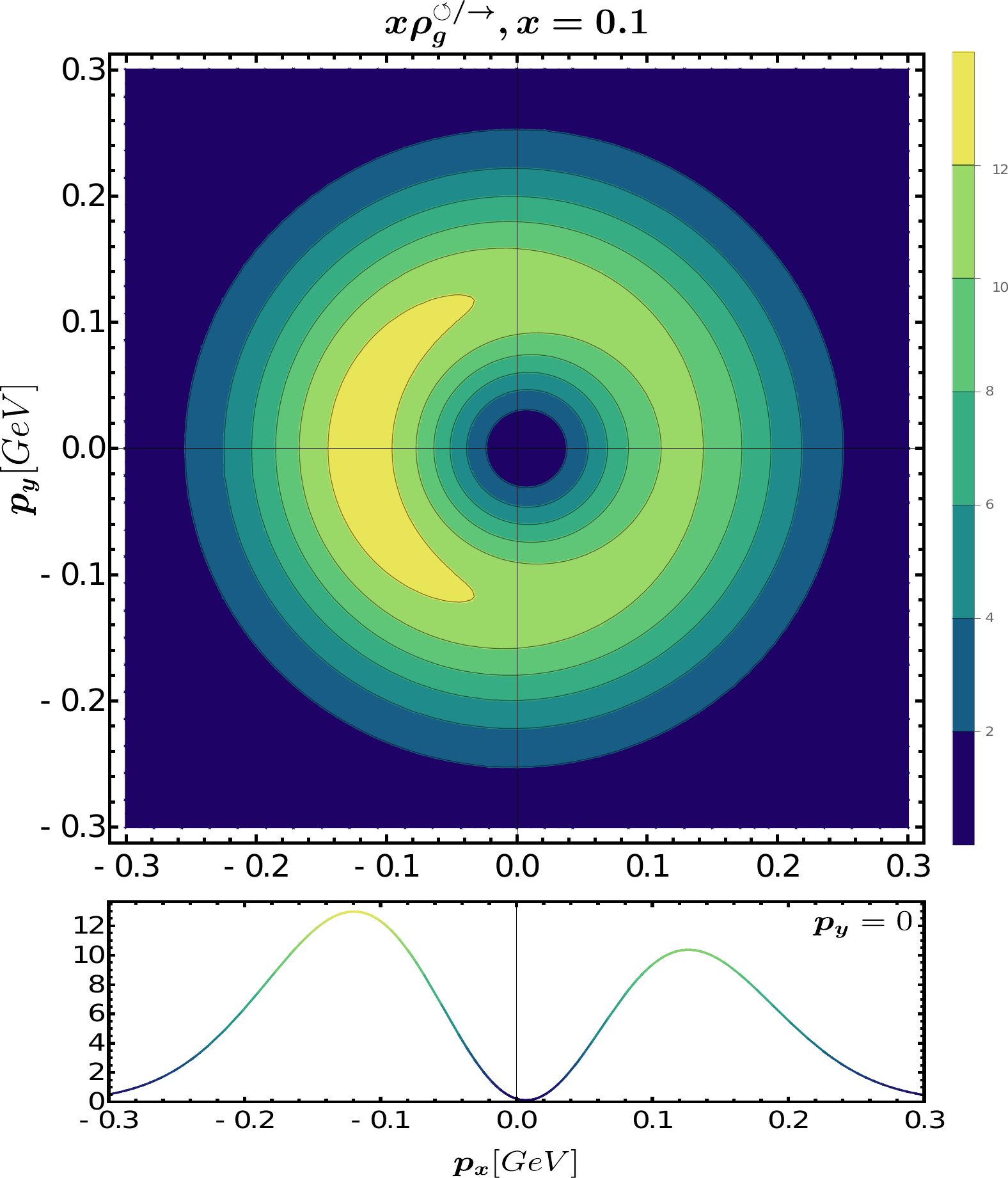}
	\caption{Upper panel: unpolarized gluon density (left), Boer-Mulders gluon density (right) for a virtually moving unpolarized nucleon.
		Lower panel: helicity gluon density (left) and worm-gear gluon density (right) for a polarized nucleon virtually moving towards the reader. 1D ancillary plots for each contour plot indicate the density at $p_y=0$.
		All densities are represented as a function of $\bfp \equiv (p_x,p_y)$ at a constant $x=0.1$.  } 
	\label{fig:densityplots}
\end{figure}
 In Fig.~\ref{fig:densityplots} we show the contour plots for the $\bfp$-distribution of the densities at $x=0.1$. The upper left panel shows the unpolarized density, $\rho_g$ which is cylindrically symmetric in the $p_{x}$ and $p_{y}$ directions followed by the ancillary 1D plots which represents the corresponding density at $p_{y} = 0$. The upper right panel represents the BM density, $\rho^{\leftrightarrow}$  which shows a quadrupole structure. 
The lower left panel presents the gluon helicity density, $\rho^{\circlearrowleft/+}$ which is perfectly symmetric in the transverse plane because it describes a proton (gluon) longitudinally (circularly) polarized along the direction of motion pointing towards the reader. The lower right panel represents the worm-gear density, $\rho^{\circlearrowleft/\leftrightarrow}$ which is slightly asymmetric in $p_{x}$ at $x=0.1$ because the proton is transversely polarized along the $p_{x}$-direction. The color code identifies the size of the oscillation of each density along the $p_{x}$ and $p_{y}$ directions.
	\subsection{ Relations between TMDs}
 The gluon TMDs are very sensitive to $x$. The TMDs and their relations among them could be separated into small and large $x$ regions. 
 Depending upon the applicability of the model these relations can be checked only in certain ranges of $x$. The leading twist TMDs in this model also satisfy the inequality relations, which are valid in QCD and all models~\cite{Lyubovitskij:2020xqj,Lyubovitskij:2021qza}, e.g., positivity bound, which is the most known model-independent relation.   According to which the unpolarized TMD $f_{1}^{g}(x,\bfp^{2})$ should be always positive and larger than the polarized one~\cite{Mulders:2000sh} i.e.,
	\begin{eqnarray}\label{eq:positivitybound1}
	f_{1}^{g}(x,\bfp^2)>0,\hspace{0.5cm}        f_{1}^{g}(x,\bfp^2)\geq|g_{1L}^{ g}(x,\bfp^2)| .
	\end{eqnarray}
 and,
 \begin{eqnarray}\label{eq:positivitybound2}
	f_{1}^{g}(x,\bfp^{2}) \geq \frac{|\bfp|}{M} |g_{1T}^{g}(x,\bfp^{2}) |,
	\hspace{0.5cm} 
f_{1}^{g}(x,\bfp^{2}) \geq 	\frac{|\bfp|^2}{2M^2}|h_{1}^{\perp g}(x,\bfp^{2})|.	
	\end{eqnarray}
 Apart from these relations, there are several relations among TMDs themselves. 
   The Mulders-Rodrigues relations for unpolarized and  
 the polarised TMDs \cite{Mulders:2000sh}  are more stringent conditions than the above positivity bounds  and are satisfied in our model:	
	\begin{eqnarray}\label{eq:inequalities}\nonumber
		\sqrt{[g_{1L}^{g}(x,\bfp^{2})]^{2}+\bigg[\frac{|\bfp|}{M}g_{1T}^{g}(x,\bfp^{2})\bigg]^{2}}\leq f_{1}^{g}(x,\bfp^{2}),\\ \nonumber
		\sqrt{[g_{1L}^{g}(x,\bfp^{2})]^{2}+\bigg[\frac{\bfp^{2}}{2M^{2}}h_{1}^{\perp g}(x,\bfp^{2})\bigg]^{2}}\leq f_{1}^{g}(x,\bfp^{2}),\\
		\sqrt{\bigg[\frac{|\bfp|}{M}g_{1T}^{g}(x,\bfp^{2})\bigg]^{2}+\bigg[\frac{\bfp^{2}}{2M^{2}}h_{1}^{\perp g}(x,\bfp^{2})\bigg]^{2}}\leq f_{1}^{g}(x,\bfp^{2}).
	\end{eqnarray}
  The positivity bounds Eqs.~(\ref{eq:positivitybound1}) and (\ref{eq:positivitybound2}) can be derived as limiting cases of Eq.(\ref{eq:inequalities}).
		
\begin{figure}
 \includegraphics[scale=0.32]
     {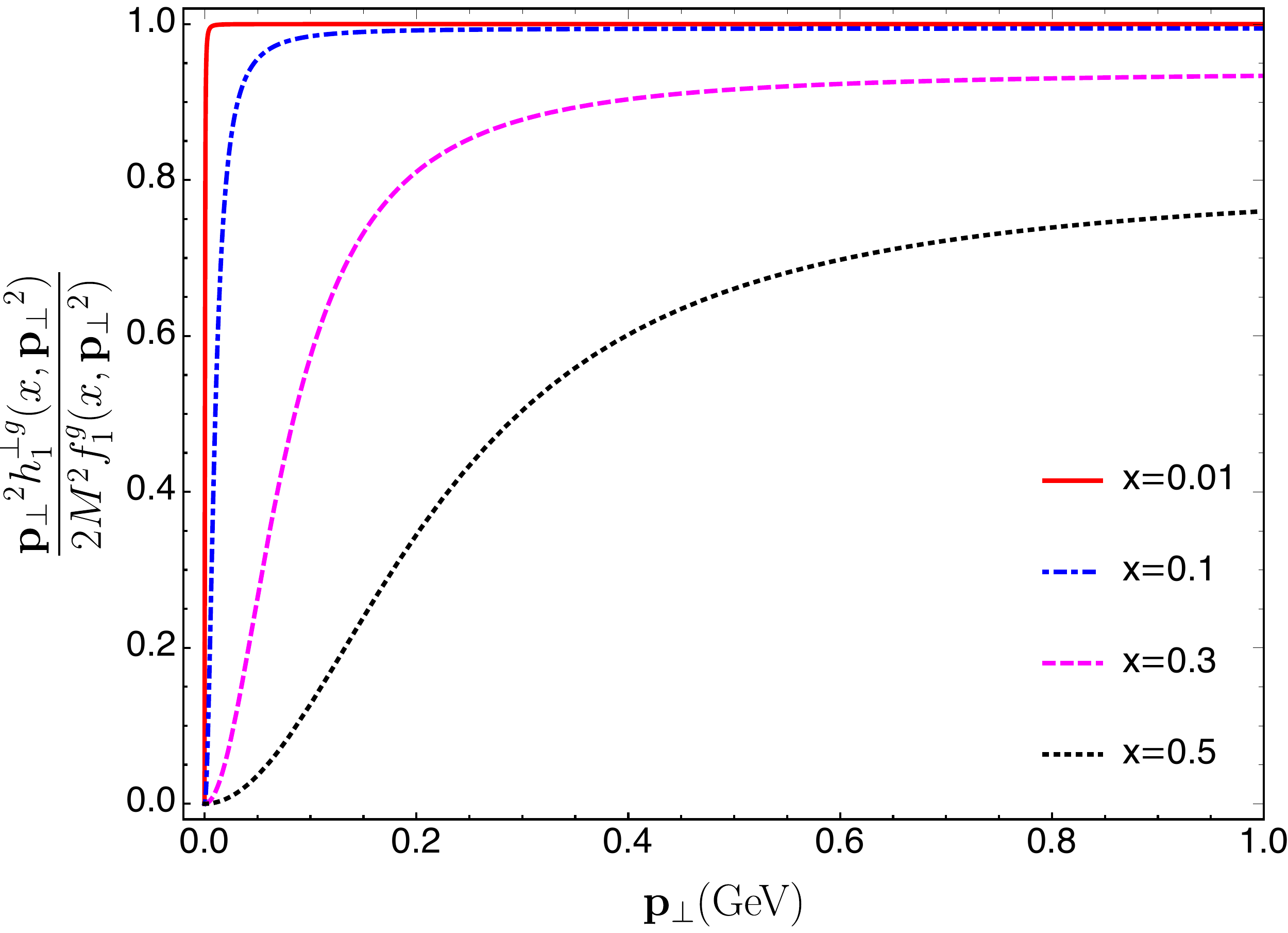}
	\caption{Positivity bound for different values of longitudinal momentum fraction $x$. The solid red curve shows the saturation of the positivity bound, whereas the other curves satisfy this constraint.}
	\label{fig:positivity}
\end{figure}

 An interesting sum rule has been derived in Ref.~\cite{Lyubovitskij:2020xqj} involving the T-even TMDs, by expressing them in terms of overlaps of LFWFs. This can be expressed as : 
	\begin{eqnarray}\label{TMDssumrule}
		[f_{1}^{g}(x,\bfp^{2})]^{2}=[g_{1L}^{g}(x,\bfp^{2})]^{2}+\bigg[\frac{|\bfp|}{M}g_{1T}^{g}(x,\bfp^{2})\bigg]^{2}+\bigg[\frac{\bfp^{2}}{2M^{2}}h_{1}^{\perp g}(x,\bfp^{2})\bigg]^{2},
	\end{eqnarray}
	The above relation, Eq.(\ref{TMDssumrule}), gives the connection between the square of an unpolarized TMD and a combination of squares of three polarised TMDs. 
In Fig.~\ref{fig:positivity}, we  show the ratio of Boer-Mulders to unpolarized TMDs weighted by $\bfp^{2}/2M^{2}$ as a function of $\bfp$ for different values of the gluon longitudinal momentum fraction $x$. 
 We notice that the positivity bound saturates only for the small $x$-values, 
for large $x$-values the positivity  inequality is  satisfied in the whole range of $\bfp$. 
 Saturation of the positivity bound for gluon TMDs in a spectator model in the certain kinematical region has been reported  in Ref.~\cite{Kishore:2022ddb}.
Note that all the relations listed above are independent of the parameters
of our model.
\section{Conclusion}\label{sec:concl}
 We  have proposed a light-front spectator model  with the light-front wave functions modelled from the soft-wall holographic AdS/QCD prediction for two-body bound states. In this simple model, proton is assumed to consist of a struck gluon and a spin-1/2 spectator. We fixed our model parameters by fitting the unpolarized gluon  PDF, $f_1^g(x)$ with the NNPDF3.0nlo global analysis. The helicity PDF and other T-even TMDs are calculated as predictions of the model and are shown to satisfy the positivity bound and have in good agreement with the available model predictions. The model is found to satisfy the constraints imposed by QCD including counting rules at small and large $x$. We  have demonstrated that the gluon TMDs obey the model-independent Mulders-Rodrigues inequalities. 
We have also shown in this model that the superposition of the squares of all polarized T-even TMDs is equal to the square of the unpolarized TMD.
We verified that this sum rule is also followed in similar models.
It will be interesting to study the other proton properties like GPDs, T-odd TMDs, Wigner distributions, GTMDs, etc., and their scale evolutions in this model, and to compare with other model predictions, which can be helpful for the upcoming EICs.

	\section*{Acknowledgements}
CM is supported by new faculty start up funding by the Institute of Modern Physics, Chinese Academy of Sciences, Grant No. E129952YR0.  CM also thanks the Chinese Academy of Sciences Presidents International Fellowship Initiative for the support via Grants No. 2021PM0023. AM would like to thank SERB MATRICS (MTR/2021/000103) for funding. The work of DC is supported by Science and Engineering Research Board under the Grant No. CRG/2019/000895.	
	 
	\bibliography{Ref.bib}

\end{document}